\documentclass[%
 reprint,
superscriptaddress,
 amsmath,amssymb,
 aps,
]{revtex4-1}

\usepackage{graphicx}
\usepackage{dcolumn}
\usepackage{bm}
\usepackage[dvipsnames]{xcolor}
\usepackage{subcaption}
\usepackage{multirow}
\usepackage[normalem]{ulem} 

\usepackage{xr}
\makeatletter
\newcommand*{\addFileDependency}[1]{
  \typeout{(#1)}
  \@addtofilelist{#1}
  \IfFileExists{#1}{}{\typeout{No file #1.}}
}
\makeatother

\newcommand*{\myexternaldocument}[1]{%
    \externaldocument{#1}%
    \addFileDependency{#1.tex}%
    \addFileDependency{#1.aux}%
}
\myexternaldocument{./SI}

\renewcommand*{\thefootnote}{\alph{footnote}}

\begin{document}


\title{Origin of Metal-Insulator Transitions in Correlated Perovskite Metals }

\author{M. Chandler Bennett}
\email{bennettcc@ornl.gov}
\affiliation{Materials Science and Technology Division, Oak Ridge National Laboratory, Oak Ridge, Tennessee 37831, USA}
\author{Guoxiang Hu}
\email{ghu@qc.cuny.edu}
\affiliation{Department of Chemistry and Biochemistry, Queens College, City University of New York, Flushing, NY 11367, USA}
\author{Guangming Wang}
\affiliation{Department of Physics, North Carolina State University, Raleigh, North Carolina 27695, USA}
\author{Olle Heinonen}
\affiliation{Materials Science Division, Argonne National Laboratory, Chicago, Illinois USA}
\author{Paul R. C. Kent}
\affiliation{Computational Sciences and Engineering Division, Oak Ridge National Laboratory, Oak Ridge, Tennessee 37831, USA}
\author{Jaron T. Krogel}
\email{krogeljt@ornl.gov}
\affiliation{Materials Science and Technology Division, Oak Ridge National Laboratory, Oak Ridge, Tennessee 37831, USA}
\email{krogeljt@ornl.gov}
\author{P. Ganesh}%
\email{ganeshp@ornl.gov}
\affiliation{Center for Nanophase Materials Sciences, Oak Ridge National Laboratory, Oak Ridge, Tennessee 37831, USA}
\email{ganeshp@ornl.gov}

\date{\today}

\begin{abstract}
The mechanisms that drive metal-to-insulator transitions (MIT) in correlated solids are not fully understood, though intricate couplings of charge, spin, orbital, and lattice degrees of freedom have been implicated. For example, the perovskite SrCoO$_3$ is a FM metal while the oxygen-deficient ($n$-doped) brownmillerite SrCoO$_{2.5}$ is an anti-ferromagnetic (AFM) insulator. Given the magnetic and structural transitions that accompany the MIT, the driving force for such a MIT transition is unclear. 
We also observe that, interestingly, the perovskite metals LaNiO$_3$, SrFeO$_3$, and SrCoO$_3$ also undergo MIT when $n$-doped via high-to-low valence compositional changes i.e. Ni$^{3+}\rightarrow$ Fe$^{4+}$, Sr$^{2+}\rightarrow$ La$^{3+}$, and Sr$^{2+}\rightarrow$ La$^{3+}$, respectively. On the other hand, pressurizing the insulating brownmillerite SrCoO$_{2.5}$ phase, drives a gap closing. Here we demonstrate that the ABO$_3$ perovskites most prone to MIT are self-hole doped materials, reminiscent of a negative charge-transfer metal, using a combination of density functional and fixed-node diffusion quantum Monte Carlo calculations. Upon $n$-doping the negative-charge transfer metallic phase, an underlying charge-lattice (or electron-phonon) coupling drives the metal to a charge and bond-disproportionated gapped insulating state, thereby achieving ligand hole passivation at certain sites only.  The size of the band gap is linearly correlated with the degree of hole-passivation at these ligand sites. Further, metalization via pressure is also stabilized by a similar increase in the ligand-hole, which in turn stabilizes the ferromagnetic coupling. These results suggest that the interaction that drives the band gap opening to realize a MIT even in correlated metals is the charge-transfer energy, while it couples with the underlying phonons to enable the transition to the insulating phase. Other orderings (magnetic, charge, orbital etc.) driven by weaker interactions may assist gap openings at low doping levels, but it is the charge-transfer energy that predominantly determines the band gap, with a negative energy preferring the metallic phase. This {\sl n}-doping can be achieved by modulations in oxygen-stoichiometry or metal-composition or pressure. Hence, controlling the amount of the ligand-hole, set by the charge-transfer energy, is the key-factor in controlling MIT.
\end{abstract}

\maketitle
\section{Introduction}
The metal-insulator transition (MIT) in various strongly correlated transition-metal oxides is essential for recently proposed device applications, for instance memristors for next-generation neuromorphic computing{\cite{zhou2015mott,andrews2019building,zhang2020understanding, Ganesh.NPJ.Nature}}. However, understanding the underlying mechanism of the MIT in correlated solids has been a longstanding problem {\cite{imada1998metal,gruner1988dynamics}}, making selection and optimization of appropriate materials difficult. The theoretical difficulty in analyzing these materials lies in the strong electron correlation, giving rise to complex many-body phenomena and a close coupling of charge, spin, orbital, and lattice degrees of freedom. Identification of the important variables and key interactions that influence the MIT in correlated metals has consequently remained elusive.

The ABO$_x$ perovskite family has several candidate compounds that undergo a sharp and tunable MIT{\cite{staub2002direct,hepting2014tunable, scherwitzl2011metal, lu2016voltage, lu2017electric}}, which forms an excellent playground to look for the driving forces behind the transition. For example, consider the oxygen-rich perovskite SrCoO$_3$ (PV-SCO) which is a correlated ferromagnetic (FM) metal. The introduction of oxygen-vacancies typically amounts to introducing negative charge-carriers (i.e. $n$-doping). In a conventional band-theory picture, such doping often increases electronic conductivity, but in PV-SCO it leads to an insulating ground-state with a concomitant change in the magnetic ordering{\cite{jeen2013reversible,choi2013reversal}}. Indeed, the oxygen-deficient brownmillerite SrCoO$_{2.5}$ (BM-SCO) is an anti-ferromagnetic insulator. However, under pressure (both uniform and uniaxial), this anti-ferromagnetic insulator shows a reduction in  the electronic band gap by ~40\%{\cite{hong2017pressure}}.  Recent experiments{\cite{chowdhury2019tuning}} suggest that under pressure, the number of holes on oxygen-ligands possibly increase, which in principle could stabilize a ferromagnetic coupling between the metal-ions, in turn eventually closing the band-gap in BM-SCO.  Given the concomitance of the MIT to structural or magnetic transitions (or both), it is not immediately clear what would be the key change that triggers the MIT in PV-SCO and other similar correlated perovskite metals. Furthermore, it is not clear whether the gap opening is driven by Hubbard repulsive interactions from correlated electrons in the 3{\sl d} orbitals, hybridization of orbitals between covalently bonded atoms, magnetic exchange interactions that favor Hund's rule, or charge-transfer energies that determine the transfer of electrons from the ligand (anion) to the metal (cation) sites. 

The family of rare-earth nickelates is another example where the MIT occurs concomitantly with structural and magnetic transitions {\cite{jaramillo2014origins, bisogni2016ground,shamblin2018experimental}}.  While the size of the rare-earth tunes the transition temperature, the underlying mechanism appears to remain the same and is dependent on the metal-ligand bonding{\cite{torrance1992systematic}}.  In all the nickelate compounds, the high-temperature phase is a correlated metal with fluctuating magnetic moments (i.e. the paramagnetic phase). Below a certain temperature, all compounds, except for LaNiO$_3$, transition into a bond-disproportionated phase, i.e. a phase with alternating long and short Ni-O bonds. Some compounds exhibit an additional transition to an anti-ferromagnetic (AFM) phase concomitant to this bond-disproportionation. For example, recent experiments suggest even LaNiO$_3$ undergoes such an AFM transition below $\sim$157K\cite{2018_Guo_Nature_Comm_9_43_antiferromagnetic_lanio}. Similarly, n-doping with hydrogen induces a metal-to-insulator transition in SmNiO$_3$, with the underlying mechanism still uncertain with the observation of local lattice distortions in the insulating phase~\cite{HSNO1,HSNO2}. Based on the Zaanen-Sawatzky-Allen (ZSA) classification scheme{\cite{zaanen1985band}}, the nickelates are self-hole doped Mott insulators where the self-hole doping is due to a negative charge-transfer energy. 

Due to a forced high-valence cationic-state (Ni$^{3+}$, $d^7$), the high-temperature metal is stabilized by self-hole doping ($d^8L$, where {\sl L} indicates a ligand-hole) and upon reducing the temperature, a {\sl bond} disproportionated insulating state emerges instead of a {\sl charge} ordered state, with short and long Ni-O bonds stabilizing alternating $d^8L^2$ and $d^8L^0$ Ni-cations, that in some cases also stabilizes an anti-ferromagnetic ground-state. 
As such, while a Hubbard repulsion is necessary, it is not sufficient to transition to an insulating phase. Indeed, even in conventionally well regarded Mott insulators, such as VO$_2$ and NiO, and their superlattices, we recently demonstrated that that charge-state of oxygen anions and its coupling to local structural distortions play a significant role in driving the MIT~\cite{VO2-paper1,VO2-paper2, VO2-paper3, VO2TiO2-paper, NiO-paper1, NiO-paper2}. Though, it is still not very clear why self hole-doping leads to a bond disproportionated phase when temperature is reduced which opens a gap, and how the ZSA theory generalizes to MIT driven by changes in pressure, composition, and stoichiometry.  

To address this question, we posit that the connection of the MIT and bond-disproportionation extends beyond the nickelate family of perovskites and is not limited to MITs triggered by a change in temperature. 
Similar to the nickelates, the ferrates and the cobaltates have metal-ions forced in high-valency as well, e.g. Fe$^{4+}$ in SrFeO$_3$ and Co$^{4+}$ in SrCoO$_3$, and thereby possess large electron affinities leading to a low, possibly negative, charge-transfer{\cite{rogge2018electronic,kunevs2012spin}} energy.
Hence, these systems should also exhibit self-hole doping to stabilize their metallic phase.  
We note that the MIT can also be triggered by compositional changes. 
Just as {\sl n}-doping PV-SCO by making it non-stoichiometric opens a band gap in the Brownmillerite phase, similarly {\sl n}-doping PV-SCO by substituting Sr$^{2+}$ for La$^{3+}$ also leads to a band gap opening. 
Similarly, LaFeO$_3$ is an insulator and can be regarded as an {\sl n}-doped PV-SrFeO$_3$ (${\rm Sr}^{2+}\rightarrow{\rm La}^{3+}$).  
Given these observations, we hypothesize that metallic ABO$_3$ perovskites that demonstrate a MIT are self-hole doped negative charge transfer metals in the Zaanen-Sawatzky-Allen (ZSA) classification scheme.
{\sl n}-doping such self-hole doped metals fills these pre-existing holes and gives rise to an insulating state, possibly via a charge-lattice (or equivalently an electron-phonon) coupling, leading to a bond-disproportionated insulating structure, that owing to a symmetry lowering transition necessarily is also charge-disproportionated (not necessarily charge-ordered). Magnetic ordering as well as other kinds of ordering, such as charge- or orbital-ordering may further assist in the gap opening of this bond-disproportionated phase. 

To prove the conjecture made above, in this work, we use density functional theory based methods together with the highly accurate fixed-node diffusion Monte Carlo (FN-DMC)\cite{GRIMM1971134} flavor of quantum Monte Carlo (QMC)\cite{RevModPhys.73.33} to quantitatively compare the degree of self-hole doping in perovskite metals, and how it correlates with changes in metal-composition, oxygen-stoichiometry and pressure, across a MIT.
As a measure of the degree of self-hole doping, i.e. $d^{n+1}L$, we compare calculations of oxygen occupations via integrated densities around the oxygen sites.
We additionally demonstrate that changes in the degree of self-hole doping triggers a charge-lattice instability.  
To benchmark our approach, we perform FN-DMC calculations of various ground-state properties and compare to experiments where possible, finding very good agreement. 
The ground-state properties used for comparison are the cohesive energies of the perovskite compounds -- indicates accuracy of describing atomic bonding -- and the local magnetization around the metal-site  which is important in determining the spin-state of the metallic cation.  
A high quality description of the ground state is foundational for our investigations of charge-transfer in the perovskite family of materials considered here. 
The fundamental band gaps from FN-DMC-benchmarked DFT$+U$ calculations are also in good agreement with available experimental gaps, with FN-DMC-gaps showing similar trends across the different compounds. 
We find that holes are present in the oxygen-site for correlated perovskite metals (as measured by the reduced oxygen occupations relative to the insulating compounds), and that transitioning to a gapped insulating phase via changes in metal-composition or oxygen-stoichiometry or pressure results in a reduction of the holes on oxygen-sites (increased oxygen occupation). This is consistent with our density of states calculations, where the metals have an unoccupied $\sl p$ band, while insulators resulting from these metals undergoing an MIT have a $\sl p$-$\sl d$ type band gaps.
We find strong evidence for the insulating phases to be bond-disproportionated, due to an underlying  charge-lattice coupling.  
We therefore prove our conjecture that self-hole doped correlated metals can trigger a metal-to-insulator transition by {\sl n}-doping into their unsaturated holes (increased oxygen occupation) due to a strong charge-lattice (or electron-phonon) coupling in perovskites, and that this {\sl n}-doping can be achieved by changes in oxygen-stoichiometry, metal-composition, or pressure.  
Hence, controlling the ligand-hole population, set by the charge-transfer energy, is the key-factor in controlling MIT, even in correlated perovskite metals with strong on-site Hubbard interactions. 

\section{Methods}
We perform a combined study using density functional theory (DFT) with a linear-response determined Hubbard $U$ (i.e. $U_{\rm
LR}$)\cite{cococcioni2005linear} as well as fixed-node diffusion Monte Carlo (FN-DMC) calculations to obtain ground-state
properties (details available in the supplementary information).  Computations were performed for metallic SrCoO$_3$ and LaNiO$_3$
compounds as well as the insulating LaFeO$_3$ and LaCrO$_3$ compounds in the perovskite phase and SrCoO$_{2.5}$ in the
brownmillerite phase, at pressures ranging from 0-8 GPa. The magnetic orderings of the experimental ground state were considered, namely ferromagnetic for
SrCoO$_3$ and LaNiO$_3$; antiferromagnetic (AFM) for LaFeO$_3$, LaCrO$_3$, LaNiO$_3$ and BM SrCoO$_{2.5}$.  For the insulating phases, we also calculated
the optical gap using LDA$+U_{\rm LR}$ and FN-DMC.  The nodal surfaces for FN-DMC were obtained by using the variational property
of FN-DMC and varying the strength of the metal-site Hubbard interaction within the LDA$+U$ to find the best nodal surface for
each material.  For the BM-SCO phase, distinct $U$ values were assumed for the two different coordination sites of the Co-atom.
For PV-SCO, we additionally used nodal surfaces obtained by varying the exact exchange fraction $w$ with PBE0$_w$. 

For this work, we utilized the \textsc{Quantum Espresso} suite\cite{Giannozzi_2009} and the \textsc{VASP}
package\cite{VASP1,VASP2,VASP3,VASP4} for the DFT$+U$ calculations and for generating the orbitals needed for QMC. All QMC
calculations were performed with the high-performance QMCPACK code\cite{Kim_2018,kent_qmcpack_2020}. Using the \textsc{Nexus}
suite\cite{KROGEL2016154}, reproducible workflows were written for the DFT calculations and the complete set of calculations
needed for QMC. 
Input  and  output  files  for  the  calculations  performed  in  this  study  are  available  and maintained at the Materials Data Facility\cite{data_doi}.

\section{Results \& Discussion}
\subsection{Implications of ground state energies and magnetic structure}
Tables \ref{tab:cohesive_energies} and \ref{tab:magnetic_moments} compare the cohesive energies and the local-moments computed from FN-DMC to available experiments, respectively.  
Cohesive energy has contributions from all necessary interactions, and is often incorrect in standard DFT-calculations
\cite{PhysRevB.84.045115}.  Here we find that the cohesive energies for LaNiO$_3$, LaFeO$_3$, and LaCrO$_3$ are in good agreement
with estimates from the literature (derivations of experimental cohesive energies is provided in the supplemental material).
Particularly, we find that the inclusion of spin-polarization is necessary to bring the cohesion of LaNiO$_3$ in closer agreement
with experimental estimates, suggesting that the physics that stabilizes the local moment is essential in describing its
electronically metallic state. The accuracy in cohesive energies for both metals and insulators indicates that FN-DMC is
appropriate for looking at the driving force behind the MIT.  In addition to cohesive energies, the methodology also captures the
magnitude of the local moments and its change with coordination and structure type for the cobaltates, with absolute values very
close to the values obtained from spin-polarized neutron scattering experiments in the
literature\cite{1957_Koehler_JPhysChemSolids_2_100_Neutron_Magnetic_Properties_LaBO3,2008_Munoz_PRB_78_054404_cryst_and_mag_structure_of_bm-sco,2018_Guo_Nature_Comm_9_43_antiferromagnetic_lanio}.
This affirms our methodology to predict the underlying electronic structure, allowing us to meaningfully conclude from systematic
changes observed in charge-/spin-densities, and derived quantities computed from the FN-DMC estimated many-body ground-state wave
function.

\begin{table}[!htbp]
\centering
	\caption{Cohesive Energies from DMC in units of eV/atom.
	}
\label{tab:cohesive_energies}
\begin{tabular}{l c c c} 
 \hline\hline
 Material    & DMC & Exp. \\
 \hline
 SrCoO$_3$        &   4.412(6)   &                                       \\ 
 LaNiO$_3$ (NM)   &   5.512(6)   & \multirow{2}{*}{5.760\footnotemark[1]} \\
 LaNiO$_3$ (FM)   &   5.656(8)   &                                       \\ 
 LaFeO$_3$        &   6.220(6)   & 6.120\footnotemark[1]                  \\ 
 LaCrO$_3$        &   6.400(6)   & 6.422\footnotemark[1]                 \\ 
 \hline
\end{tabular}
\footnotetext[1]{See supplemental material for derivations of experimental values.}
\end{table}

\begin{table}[!htbp]
\centering
	\caption{Local Magnetic moments (Bohr mag) of cations from FN-DMC. The local moments are calculated by spherically averaging the $\rho_{\rm up}-\rho_{\rm down}$ density (where $\rho_{\rm up/down}$ is the spin-up/spin-down density) around the cation site.
	}
\label{tab:magnetic_moments}
\begin{tabular}{l l c c c} 
 \hline\hline
 Material      & Atom & DMC & Exp.                      \\
 \hline
 SrCoO$_3$        &   Co & 2.563(1)   & 2.5                       \\ 
 LaNiO$_3$(FM)    &   Ni & 1.019(1)   & \multirow{2}{*}{0.3 \footnotemark[1]} \\ 
 LaNiO$_3$(G-AFM) &   Ni & 0.518(2)   &                            \\
 SrCoO$_{2.5}$    &  Co1 & 3.187(1)   & 3.12(13) \footnotemark[2] \\ 
 SrCoO$_{2.5}$    &  Co2 & 3.081(1)   & 2.88(14) \footnotemark[2] \\ 
 LaFeO$_3$        &   Fe & 4.202(1)   & 4.6(2)   \footnotemark[3] \\ 
 LaCrO$_3$        &   Cr & 2.537(2)   & 2.8(2)   \footnotemark[3] \\ 
 \hline
\end{tabular}
\footnotetext[1]{Reference \cite{2018_Guo_Nature_Comm_9_43_antiferromagnetic_lanio}}
\footnotetext[2]{Reference \cite{2008_Munoz_PRB_78_054404_cryst_and_mag_structure_of_bm-sco}}
\footnotetext[3]{Reference \cite{1957_Koehler_JPhysChemSolids_2_100_Neutron_Magnetic_Properties_LaBO3}}
\end{table}

Rhombohedral LaNiO$_3$ was thought to be a paramagnet, until very recent experiments on single-crystals of LaNiO$_3$ demonstrates
that it undergoes a transition to an anti-ferromagnetic phase below $\sim$157K, while still remaining
metallic\cite{2018_Guo_Nature_Comm_9_43_antiferromagnetic_lanio}.  Nevertheless, the measured local moment was $\sim$0.3$\mu_B$, much smaller than the
moments on other perovskite systems, in good qualitative agreement with trends observed from our FN-DMC calculations, which even
in the ferromagnetic phase of rhombohedral LaNiO$_3$ finds a lower moment of $\sim$1.02$\mu_B$  (Table \ref{tab:magnetic_moments})
compared to other perovskite magnets. Note that our FN-DMC moment for the G-type AFM ordering of LaNiO$_3$ was also small
($\sim$0.52$\mu_B$), however, this ordering was not energetically favored in FN-DMC over FM ordering. 
In all cases, we have used spin symmetry-broken trial wavefunctions in FN-DMC, which may partially explain 
the larger fluctuations observed when the predicted moments are small in magnitude.
The local moments of all other
compounds from our fully {\sl ab initio} DMC calculations, including those on the different Wyckoff positions for the
Brownmillerite phase, show excellent agreement with experiments. 

\begin{figure*}[t]
     \centering
     \begin{subfigure}[b]{0.45\textwidth}
         \centering
         \includegraphics[width=\textwidth]{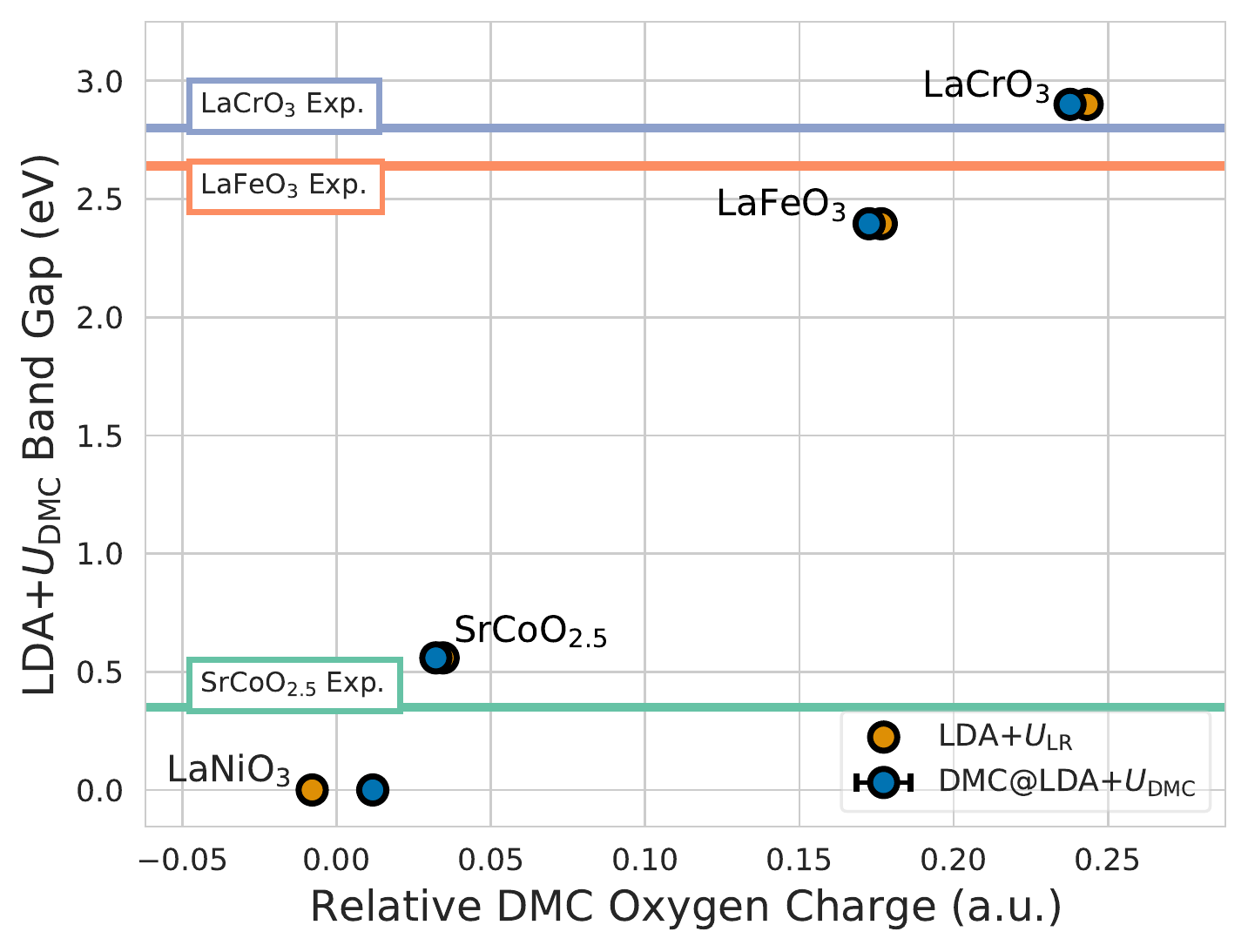}
         \caption{}
         \label{fig:gaps_vs_charge}
     \end{subfigure}
     \hfill
     \begin{subfigure}[b]{0.45\textwidth}
         \centering
         \includegraphics[width=\textwidth]{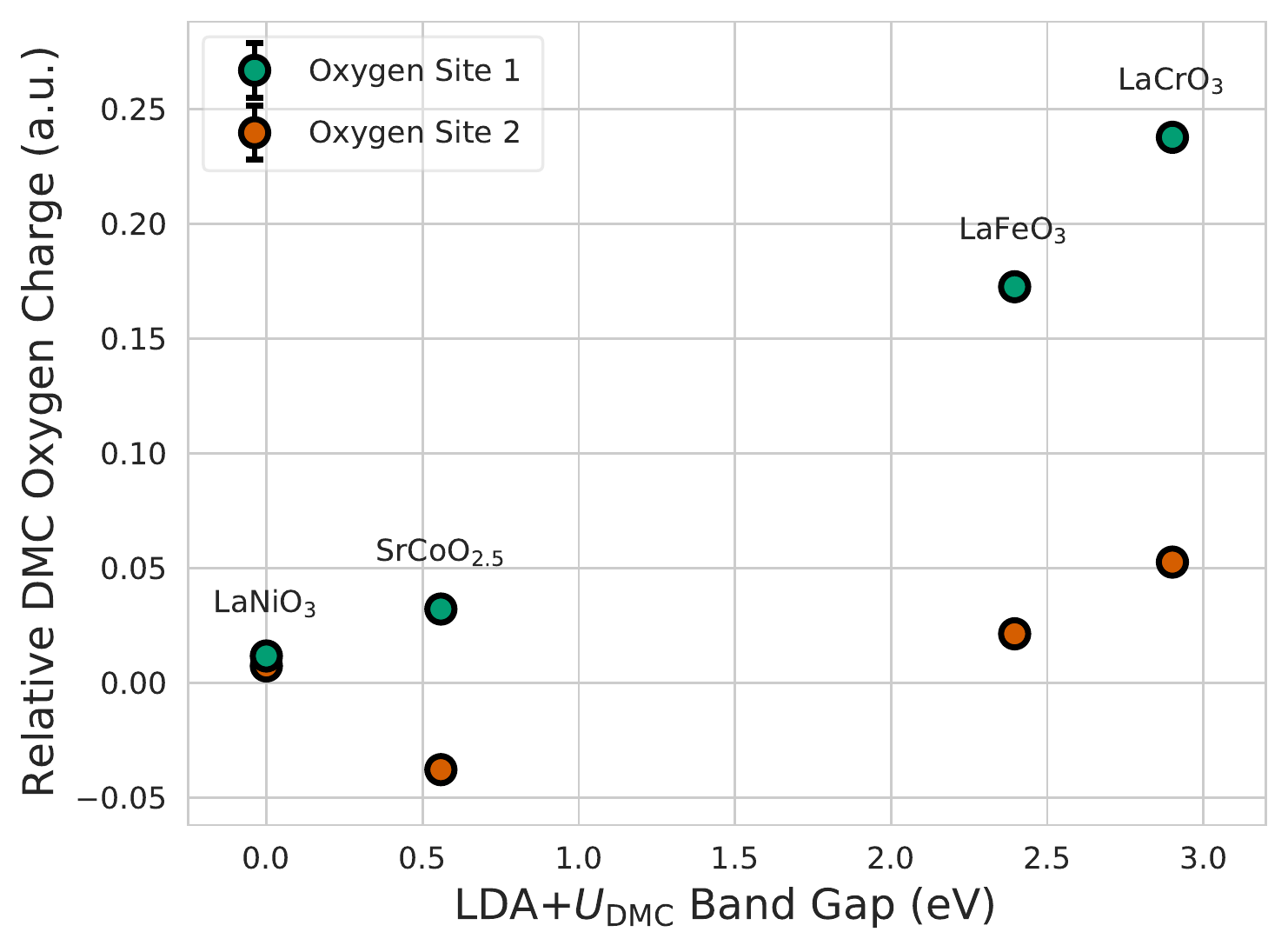}
         \caption{}
         \label{fig:charge_vs_gaps}
     \end{subfigure}
     \caption{(a) For LaNiO$_3$, SrCoO$_{2.5}$, LaFeO$_3$, and LaCrO$_3$, optical band gaps calculated within LDA$+U$ versus the DMC cumulative radially-averaged charge around the oxygen site relative to SrCoO$_3$ (with O-charge 6.4120 a.u.). The oxygen site with majority charge was used for each system and the value of $U$ that minimizes the FN-DMC energy is chosen for each system. Cumulative charges are also shown from LDA$+U$ with $U$ values consistent with DMC. For comparison, we indicate experimentally measured band gaps for the insulating systems (SrCoO$_{2.5}$\cite{PhysRevLett.111.097401}, LaFeO$_3$\cite{lfo_exp_gap}, and LaCrO$_3$\cite{lcro_exp_gap}) with horizontal solid lines. (b) LDA$+U_{\rm DMC}$ band gap versus site-specific cumulative radially-averaged charge around the oxygen site relative to SrCoO$_3$. 
     }
     \label{fig:charge_gap}
\end{figure*}

The on-site moments may be understood qualitatively in the atomic limit as $\sim 2 \sqrt {s(s+1)}$, where $s$ is the spin multiplicity, we find that the metal-sites under an octahedral crystal-field splitting are close to a high-spin configuration for LaCrO$_3$ ($t_{2g}^3e_g^0$)) and LaFeO$_3$ ($t_{2g}^3e_g^2$), an intermediate spin state for SrCoO$_3$ ($t_{2g}^4e_g^1$) and a low-spin state for LaNiO$_3$ ($t_{2g}^6e_g^1$).  Interestingly, both linear-response theory as well as FN-DMC orbital optimization predicts an increasing trend in the strength of the on-site Hubbard repulsion for 3{\sl d} electrons as we increase the {\sl d}-orbital filling from LaCrO$_3$ to LaFeO$_3$,  but it is reduced slightly as we move to SCO and LNO (details in SI).  But the variation overall is not large and both the $U_{\rm LR}$ and the $U_{\rm DMC}$ remain between 4-5.5 eV. Further, the band gaps do not follow the trend of the Hubbard interaction, with LaCrO$_3$ having a larger gap than LFO, and the SCO with similar value for the Hubbard interaction as LaCrO$_3$ showing a metallic phase. This suggests that while the Hubbard interaction parameter (including the magnetic exchange, which dictates the Hund's rule) is necessary to stabilize the spin-state, and determine the local moment, the opening  of the band gap is determined in large measure by other interactions, such as charge-transfer.  Further, the responsible interaction necessarily also destabilizes the high-spin state favored by Hund's rule in both SCO and LNO.  Because all compounds have the same oxygen ligand, one way to assess the degree and nature of charge-transfer across these compounds is to compare the charge associated with it. 

\subsection{Understanding key ligand charge behavior}
To understand what type of charge-transfer could be present in our systems, we next compare the charges on the oxygen ligand sites. The charges were computed by integrating the radial charge-density around oxygen atoms within a fixed radius for all compounds, from both the linear-response LDA$+U_{\rm LR}$ as well as the FN-DMC following a nodal optimization using LDA$+U$ orbitals.  The computed charges are in good agreement between the different methods (see SI Tables III-V), suggesting that the linear-response captures the bonding characters in the solid appreciably well. As also seen in Figure \ref{fig:charge_gap}, the LDA$+U_{\rm LR}$ band gaps are  in decent agreement with the experimental gaps.  A high degree of charge-transfer from oxygen-ligands to the metal-site is indicative of a negative charge-transfer energy.  This is what we observe from Fig.\ref{fig:charge_gap}a and \ref{fig:charge_gap}b, where the metallic SCO and LNO are seen to have lowest relative charges on the oxygen atoms compared to the mean or majority charges carried by the oxygen ligand amongst all the compounds.  Interestingly, the change in the majority charge on the oxygen ligand correlates linearly with the band gap, irrespective of whether we consider the observed experimental gap, or the band gaps calculated via LDA$+U_{\rm LR}$ or LDA$+U_{\rm DMC}$, and in spite of the change in the structure from perovskite to a brownmillerite phase.  This clearly suggests that the quantity determining the size of the band gap across MIT in correlated perovskites is the charge-transfer energy, and that the metals are self-hole doped because it is negative.  

A negative charge-transfer energy will also lead to a change in the formal oxidation state of the metal site, and thereby the magnetic moment. For example, in PV-SCO, the nominal Co$^{4+}$ ($3d^5$) valency will get lowered due to a transfer of electron from the ligand to the metal-site, destabilizing the nominally high-spin $t_{2g}^3e_g^2$ state, and giving rise to an intermediate spin-state $t_{2g}^4e_g^1$ with a net moment less than $\sim$3$\mu_B$, which would be in good agreement with both experiments and our calculations (Table \ref{tab:magnetic_moments}).  Our integrated charges around the Co-site also supports this conclusion, with the nominal valency of $\sim$2.65a.u., closer to a Co$^{3+}$ as shown in Table SI-\ref{table_SI:qmc_metal_charges}.  Similarly, hole-doping is seen to lower the nominal Ni$^{3+}$ to be $\sim$1.73a.u., closer to Ni$^{2+}$.  With a reduced valency, it would be possible to stabilize a low-spin state, such as $t_{2g}^6e_g^2$, explaining the low moments in both the measured and our computed values.  Note that this would also require the holes to have an $e_g$ symmetry so as to quench the high-spin moments in the $e_g$ sub-bands of the Co and Ni-atoms. As we will see below, this is just what we observe.  

While the nominal charges are much reduced for the metallic PV-SCO and PV-LNO, consistent with a negative charge-transfer picture, the insulating perovskites have a charge-state closer to the nominal valency e.g  a valency of $3+$ in LaFeO$_3$ (2.50 a.u.) and LaCrO$_3$ (2.78 a.u.) (Table SI-\ref{table_SI:qmc_metal_charges}).  Nominally, an isolated oxygen atom possesses 6 a.u. of charge.  We note that our pseudopotentials use a 2-electron core for oxygen and a 10-electron core for the 3$d$ metals. Due to formation of metal-ligand bonding, and owing to its high electronegativity, oxygen atoms are seen to possess more than 6 a.u. of charge for the compounds.  Remarkably, the amount of oxygen-ligand charge appears to be the same (6.41 a.u.) for the two metallic systems -- PV-SCO and PV-LNO, even though the total charges on the metal-sites (Co \& Ni) are very different (14.35 \& 16.27 a.u.) (see Table SI-\ref{table_SI:scf_metal_charges}).  This again underscores the importance of the anion O-$p$ states in determining the overall electronic-structure. Magnetic transitions could in certain cases result from a MIT,  such as in PV-SCO -- but do not primarily drive it.
Indeed, even LaNiO$_3$ was found to be a metal, a rarity for a material with an AFM ground state, consistent with experiments. This difference relative to broadly observed FM correlated metals suggests that modified magnetic interactions that could result in a particular type of AFM ordering for LaNiO$_3$ and FM for other materials may not be the primary determinant for gap opening in self hole-doped {\bf perovskite} metals.
Given that the gap varies monotonically with the formal electronegativity of the metal-atom (see Fig. SI-\ref{fig_SI:FigX_GapVsElecNeg}), it is clear that the opening of the gap in these self hole-doped metals would require changing the nature of this charge-transfer energy, i.e making it more positive as the gap increases.  

To investigate how the nature of the gap changes, we look at the orbitals participating in the gap opening. If there is an energy cost to transfer charges from the O-$p$ orbitals to the metal-$d$ orbitals, then the system would be gapped.  Such a gap is naturally realized when the Fermi-level moves from the middle of the O-$p$ band, such as the case in the self-hole doped metals, to the top of the O-$p$-band, giving rise to a $p-d$ type gap. A larger Hubbard interaction would naturally yield a larger $p-d$ gap. Indeed, as shown in the calculated projected density-of-states, Fig. SI-\ref{fig_SI:FigS1_DOS} in SI, both metallic SCO and LNO show a relatively large number of states with an O-$p$ character at the Fermi-level. As we move out of this metallic phase by reducing the amount of charge-transfer from O-$p$ to the metallic $d$ state, by stoichiometrically or compositionally $n$-doping it, a gap is opened in the insulators BM-SCO (Fig. \ref{fig:bm_sco_enthalpy_dos}) and LaFeO$_3$ (Fig. SI-\ref{fig_SI:FigS1_DOS}). These compounds show a $p-d$ type band gap, reminiscent of a more positive charge-transfer system. As we further move to LaCrO$_3$, the amount of charge on the ligand site further increases, with the Cr-metal showing a more formal oxidation state of $+3$, as noted above.  At this point the band-gap is equally determined by the Hubbard interaction and the charge-transfer energy, as also evidenced in the increased $d$ contribution to the valence-state in the projected density-of-states (Fig. SI-\ref{fig_SI:FigS1_DOS}).  

\begin{figure*}[t]
\centering
\includegraphics[width=7in]{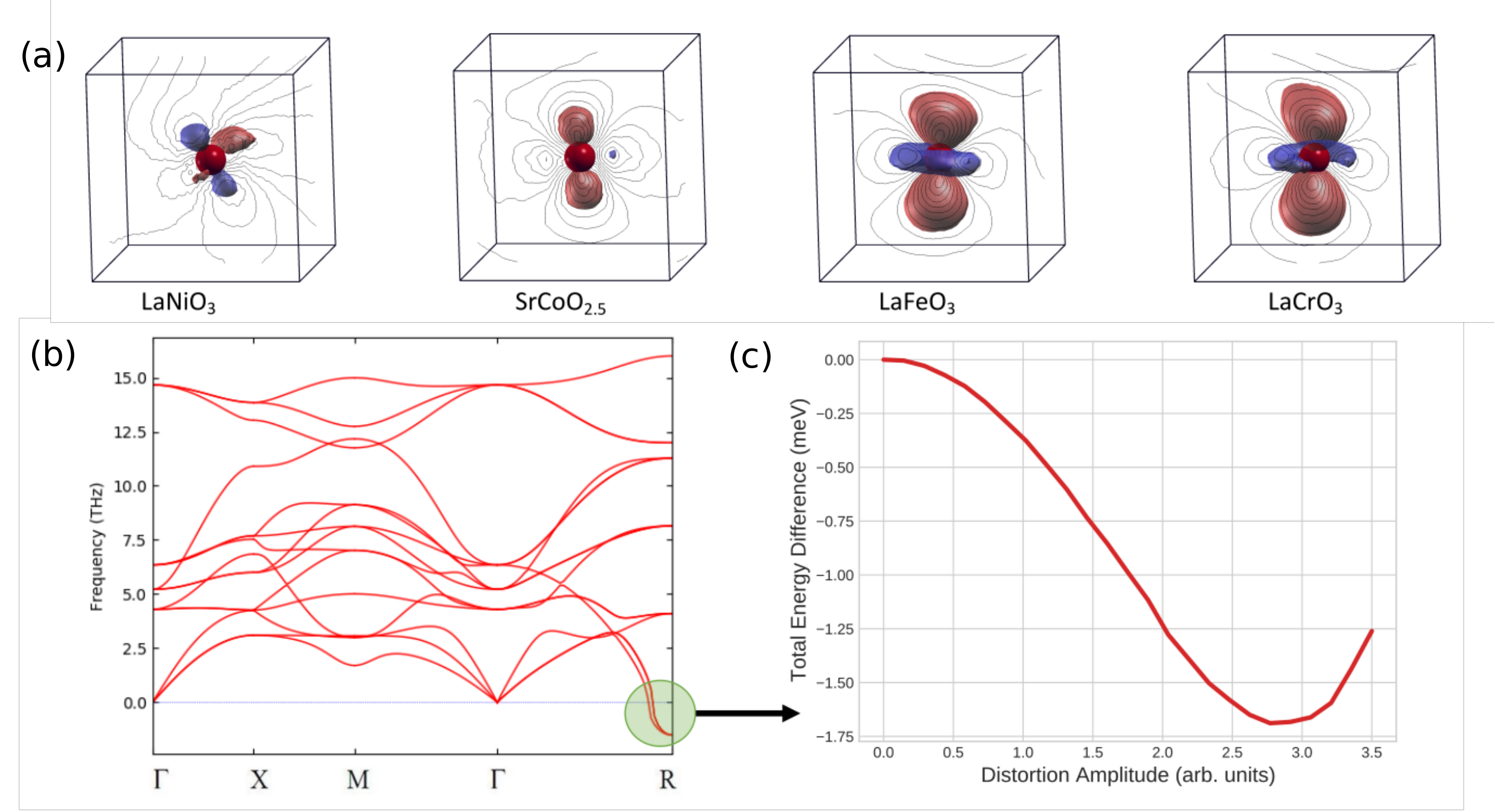}
\caption{(a) DMC charge-density isosurface differences for the indicated compounds with respect to PV-SCO plotted around an apical O-atom.  An increase in charge is shown in `red' while a decrease in `blue'. For each of the isosurfaces, the oxygen atom is centered in a cube of length 1.4 \AA and we take an isovalue of 0.022 a.u. so that each graphic can be soundly compared. (b) Phonon bands for $e^-$-doped PV-SCO showing dynamic instability at the $R$-point corresponding to an octahedral rotation. (c) Energy-surface along the distortion corresponding to the instability at the $R$-point suggesting a tendency for bond-disproportionation due to charge-disproportionation.}
\label{fig:phonon_chacomparedrge_iso}
\end{figure*}

The charge-difference with respect to a self-hole doped metal indicates how the density would respond to electron doping such a metal, and represents the nature of the lowest excitation. This is clearly seen when we plot the DMC charge-density isosurfaces, localized around the oxygen sites as shown in Fig. \ref{fig:phonon_chacomparedrge_iso}a for the different compounds, in reference to the metallic PV-SCO.  While the difference between metallic LNO and SCO shows the hole to have more of a $t_{2g}$-symmetry, the gapped insulators show the hole state to have more of an $e_g$ symmetry.  This is exactly what we expect, because for the self-hole doping to destabilize the high-spin states in the metallic SCO and LNO, as discussed above, the states involved in the electron-transfer need to have an $e_g$ symmetry.  Further, with increasing {\sl n}-doping the magnitude of this charge-transfer increases, suggesting that the charge-transfer energy is becoming more positive.  Thinking of this charge-density difference as the lowest type of excitation from the sea of electrons in the self hole-doped metal to form an $n$-doped insulator, we see that the nature of this excitation is $p-d$-like, in agreement with the predominantly  $p-d$ like gaps we observe for all systems in their density-of-states (Fig. SI-\ref{fig_SI:FigS1_DOS}).  These observations suggest that {\sl n}-doping a negative charge-transfer metal can result in making the charge-transfer energy more positive. However, it is not clear if $n$-doping simply shifts the Fermi-level to open a band gap when the Fermi-level reaches the top of the $p$-band or if there is some kind of instability that opens a gap when the system is perturbed away from its self-doped metallic state. 

Notice that while the majority charge on the oxygen ligand showed a monotonic increase with {\sl n}-doping the self hole-doped metal, the compounds show a symmetry lowering transition, which we can also associate with {\sl charge}-disproportionation.  Indeed, while one of the oxygen-ligand site shows an increase in charge (Fig. \ref{fig:charge_vs_gaps}), the others have the same electron count as in the metallic phase.  Also, a {\sl shorter} metal-ligand bond length is associated with an oxygen site with {\sl less} charge in all the compounds we have investigated, as shown in Table SI-\ref{table_SI:bond_lengths_and_charges}. This suggests a clear connection between charge and bond disproportionation.  Interestingly enough, as the band-gap increases, the amplitude of the charge disproportionation also increases.  Without further information, it is not immediately clear if the charge disproportion is caused by the bond disproportionation or vice versa, and what if any is its role in opening the band gap.  But it is clear that under $n$-doping when a self-hole doped metal undergoes an insulating transition, symmetry is always lowered. 

\subsection{Charge-lattice coupling in SrCoO$_3$}
To further understand this connection between charge and bond disproportionation, and explain the symmetry lowering, we subjected
cubic SrCoO$_3$ to the same level of electron doping (~0.25 $e^-$/Co) as the largest radial charge transfer we observed in our
study.  Interestingly, as shown in Fig.~\ref{fig:phonon_chacomparedrge_iso}b, $n$-doped PV-SCO shows a phonon-instability.  The
instability is at an $R$ point of the Brillouin zone,  corresponding to an octahedral rotation. Indeed, a similar $R$ type
octahedral rotation underlies the ground-state structure of LaFeO$_3$, leading to its rhombohedral symmetry. Freezing the
octahedral rotation lowers the energy, as seen in Fig. \ref{fig:phonon_chacomparedrge_iso}c, suggesting a weak, but persistent
charge-lattice (i.e. electron-phonon) coupling.  Similarly, hole-doping LaFeO$_3$ also gives rise to a phonon instability (Fig.
SI-\ref{fig_SI:FigY_HoleDopedPhonon_LFO}). This suggests that as the correlated metal becomes $n$-doped, either via removing
oxygen-atoms or chemical substitution (or by applying strain as discussed below), the system undergoes bond disproportionation due
to the underlying charge-lattice coupling, which in turn can lead to a charge disproportionated state as the symmetry gets
lowered. Indeed, even in hole-doped correlated metals that show a charge-ordered ground state, phonon damping and changes in heat
conduction have been experimentally observed, supporting this thesis \cite{Cohn2000}.  Such bond and charge disproportionation can
remove electronic degeneracies and stabilize other symmetry breaking transitions such as magnetic, charge-ordering or
orbital-ordering, eventually opening a band gap.   

\begin{figure*}[!ht]
\centering
\includegraphics[width=\textwidth]{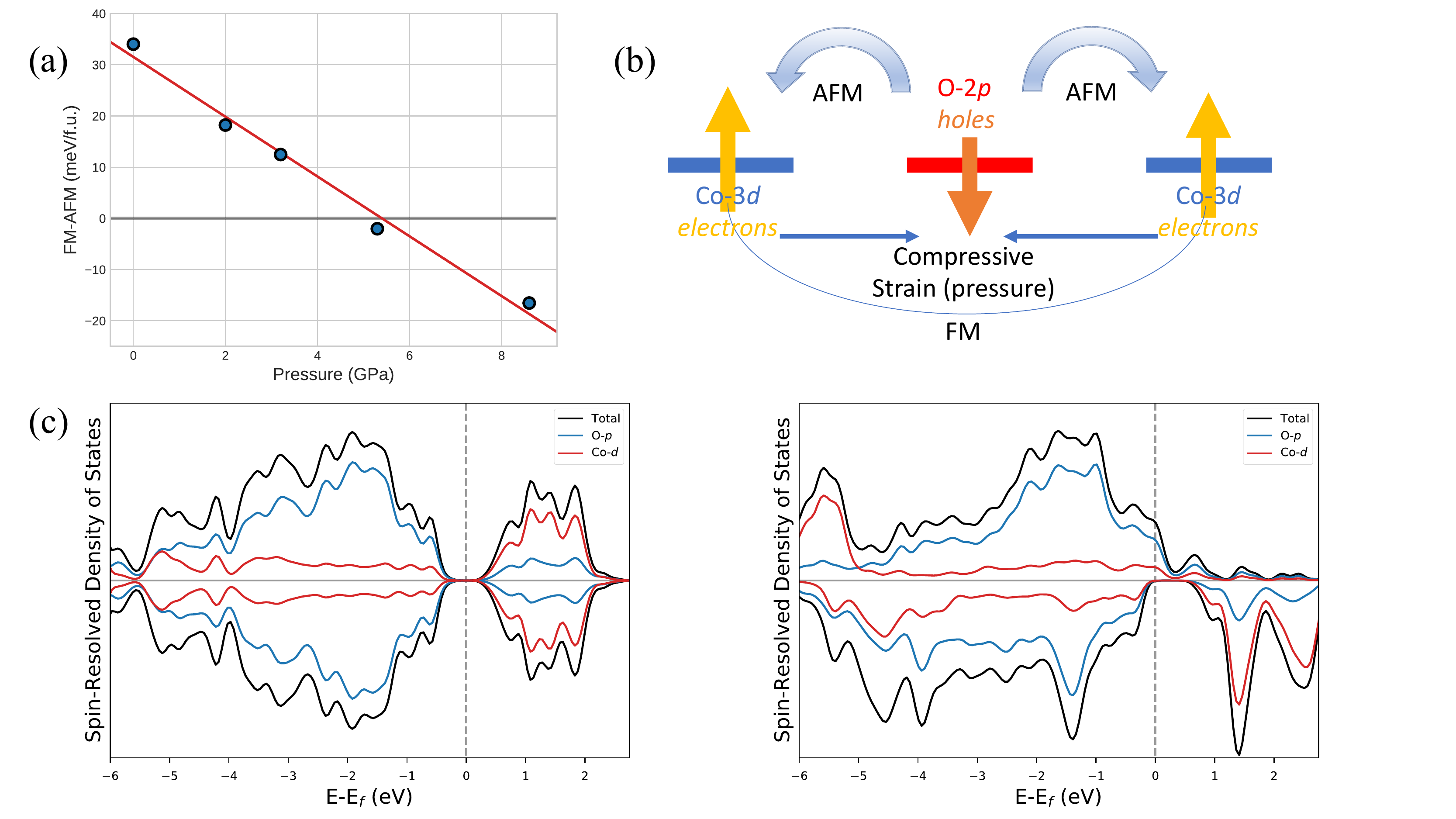}
\caption{(a) LDA$+U$ enthalpy difference in BM-SCO as a function of strain. (b) Scheme of how holes mediate the magnetic transition. (c) LDA$+U$ DOS plots of AFM insulator and FM metal phases of BM-SCO. }
\label{fig:bm_sco_enthalpy_dos}
\end{figure*}
While we demonstrated how a self-hole doped metal can become an insulator due to change in metal-composition and oxygen-stoichiometry, pressure can achieve the same result.  Indeed, a recent experiment suggests that pressurizing the antiferromagnetic BM-SCO phase leads to a reduction in the band-gap {\cite{hong2017pressure}}.  While beyond 8.5 GPa the system appeared to transform into a different structure, up to 8.5 GPa a monotonic reduction in the optical gap was seen with pressure without any structural change. Indeed, BM-SCO is an insulator in the antiferromagnetic phase but a metal in the ferromagnetic phase, as seen in our density-of-states plot in Fig. \ref{fig:bm_sco_enthalpy_dos}c.  By plotting the enthalpy difference between the two magnetic orderings as a function of strain induced by pressure (Fig. \ref{fig:bm_sco_enthalpy_dos}a), we find that the FM-ordering becomes more stable at higher pressures.  Further investigating the amount of holes on the O-$p$ orbitals, we find that with pressure the amount of charge on the oxygen ligand sites decrease monotonically. The charges were computed within the same radius, scaled to the reduction in volume, shown in Fig. SI-\ref{fig_SI:bm_sco_strained_o_populations}.
A decrease in the charges indicate an effective increase in the amount of O-$p$ holes.
Presence of such holes can stabilize the ferromagnetic coupling between the Co-sites, as shown in Fig. \ref{fig:bm_sco_enthalpy_dos}b.
Indeed, small local moments observed on the O-$p$ sites (0.17 $\mu_B$) had an opposite orientation to the ferromagnetically coupled moments on the Co-site.  This clearly suggests, that the reduction in the band gap and its eventual closing under the application of pressure in BM-SCO insulator is also driven by the same charge-transfer energy, with compressive (tensile) strain acting as $p$ ($n$)-doping of the system which thereby modifies the ligand holes, and that the AFM ordering in unstrained BM-SCO is a secondary effect that further assists in stabilizing the gap. Our findings also elucidate the physics underpinning recently observed machine-learning based identification of the average deviation of covalent radii and the global instability index as features that predominantly correlate with the tendency of materials to undergo a thermally driven MIT~\cite{ML_Rondinelli}.

\section{\label{sec:level4}Conclusions} In this work, we use density functional theory based methods and the diffusion Monte Carlo
(DMC) flavor of the many-body quantum Monte Carlo (QMC) approach that explicitly treats strong electron-electron correlations, to
understand what drives MIT in correlated perovskites. Where experimental data is available, comparisons  cohesive energies, local
moments and other quantities are made. We find the correlated metal such as PV-SCO and PV-LNO to be self-hole doped as measured by significant O$p$ contributions to conduction bands and decreased oxygen occupation relative to the studied insulators, with similar
amounts of ligand-hole. The electronic band gap in insulating compounds that are related to these self-hole doped metals via {\sl
n}-doping, due to changes in oxygen-stoichiometry such as SrCoO$_{2.5}$, or metal-compositions, such as, LaFeO$_3$ and LaCrO$_3$, show filling of this ligand-hole in some sites, with the gap nearly
linearly changing with the degree of average (or majority) filling, suggesting that the band gap is opened by charge-transfer
energies. Further, the insulating phases are shown to be {\sl both} charge and bond-disproportionation due to an underlying charge-lattice coupling,
with more ligand-holes on certain oxygen-sites than others in the symmetry lowered bond-disproportionated phase. The charge-disproportionation drives bond-disproportionation, as is seen from phonon instabilities that arise when $n$-doping a pristine cubic perovskite phase.   Together, this leads to the opening of a band gap, that is more {\sl p-d} like beyond a critical amount of {\sl n}-doping, transitioning of a self-hole doped metal to a less self-hole doped insulator.  Similarly, we find that pressure leads to an increase in ligand holes
in the oxygen-sites, driving a transition to a ferromagnetic metallic ground-state.  

Our study thus suggests that self-hole doped correlated metals can trigger a metal-to-insulator transition by {\sl n}-doping into
a less self-hole doped insulator as doping makes the charge-transfer energy more positive, and that this can lead to symmetry lowering
transitions due to a strong charge-lattice (or electron-phonon) coupling. This {\sl n}-doping can be achieved by
modulations in oxygen-stoichiometry or metal-composition or pressure. Moreover, this tendency to remain self-hole doped in correlated metals determines a universal electronic response to modulations in oxygen-stoichiometry/metal-composition/pressure via the charge-lattice coupling.  Hence, controlling the amount and anisotropy of the ligand-hole is the
key-factor in controlling MIT even in correlated metals, and the band gap is fundamentally controlled by the strength of the
charge-transfer energy close to the MIT, and not by the Mott-Hubbard interactions as originally thought for Mottronics
applications \cite{Chen2020}. 
While we do not present a rigorously predictive quantitative model that explains all of our observations, we have shown clearly that there exists a linear trend between the calculated band gaps and oxygen occupations across changes in composition, stoichiometry, or pressure. In addition, one could argue that the same trend would be seen for the hybridization between the transition metal and ligand site, which is related to the change in occupation, however hybridization cannot be quantitatively measured in the solid state so we use occupations as a proxy.
This knowledge can be used to discover new materials where MIT can be controlled more reliably enabling low-power and efficient neuromorphic devices\cite{Ganesh.NPJ.Nature}. 

\section*{Acknowledgements}
This work was supported by the U.S. Department of Energy, Office of Science, Basic Energy Sciences, Materials Sciences and
Engineering Division, as part of the Computational Materials Sciences Program and Center for Predictive Simulation of Functional
Materials. Part of this research was conducted at the Center for Nanophase Materials Sciences (CNMS), which is a DOE Office of Science User Facility. This research used resources of the Oak Ridge Leadership Computing Facility at the Oak Ridge National Laboratory, which
is supported by the Office of Science of the U.S. Department of Energy under Contract No. DE-AC05-00OR22725. This research used
resources of the National Energy Research Scientific Computing Center, a DOE Office of Science User Facility supported by the
Office of Science of the U.S. Department of Energy under Contract No. DE-AC02-05CH11231. This manuscript has been authored by
UT-Battelle, LLC under Contract No. DE-AC05-00OR22725 with the U.S. Department of Energy. The United States Government retains and
the publisher, by accepting the article for publication, acknowledges that the United States Government retains a nonexclusive,
paid-up, irrevocable, worldwide license to publish or reproduce the published form of this manuscript, or allow others to do so,
for United States Government purposes. The Department of Energy will provide public access to these results of federally sponsored
research in accordance with the DOE Public Access Plan (http://energy.gov/downloads/doe-public-access-plan).

C.B. and G.H. contributed equally to this work.  

\bibliography{abox}
\end{document}


\preprint{APS/123-QED}

\title{Origin of Metal-Insulator Transitions in Correlated Perovskite Metals}
\author{M. Chandler Bennett}
\email{bennettcc@ornl.gov}
\affiliation{Materials Science and Technology Division, Oak Ridge National Laboratory, Oak Ridge, Tennessee 37831, USA}
\author{Guoxiang Hu}
\email{ghu@qc.cuny.edu}
\affiliation{Department of Chemistry and Biochemistry, Queens College, City University of New York, Flushing, NY 11367, USA}
\author{Guangming Wang}
\affiliation{Department of Physics, North Carolina State University, Raleigh, North Carolina 27695, USA}
\author{Olle Heinonen}
\affiliation{Materials Science Division, Argonne National Laboratory, Chicago, Illinois USA}
\author{Paul R. C. Kent}
\affiliation{Computational Sciences and Engineering Division, Oak Ridge National Laboratory, Oak Ridge, Tennessee 37831, USA}
\author{Jaron T. Krogel}
\email{krogeljt@ornl.gov}
\affiliation{Materials Science and Technology Division, Oak Ridge National Laboratory, Oak Ridge, Tennessee 37831, USA}
\email{krogeljt@ornl.gov}
\author{P. Ganesh}%
\email{ganeshp@ornl.gov}
\affiliation{Center for Nanophase Materials Sciences, Oak Ridge National Laboratory, Oak Ridge, Tennessee 37831, USA}
\email{ganeshp@ornl.gov}

\date{\today}

{\bf Supplementary Information:}

\maketitle

\section{\label{sec:level2}Methods}

\subsection{\label{subsec:level1}DFT+U}

Density functional theory (DFT) calculations were performed using the Quantum Espresso (QE) software package. The GBRV pseudopotentials and local density approximation (LDA) exchange correlations (XC) were used. The plane wave cutoff was set to 120 Ry for the wave function. Energies were converged with a $1 \times 10^{-7}$ Ry tolerance. The experimental structures of the perovskites from Inorganic Crystal Structure Database (ICSD) were used for the calculations. The primitive cell of SrCoO$_3$, LaNiO$_3$, LaFeO$_3$, and LaCrO$_3$ contains 5, 10, 20, and 20 atoms, respectively. An $8 \times  8 \times 8$ k-point grid was used for the 5-atom and 10-tom cells, while an $8 \times 8 \times 5$ k-point grid was used for the 20-atom cell. The primitive cell of the oxygen deficient brownmillerite (BM) SrCoO$_{2.5}$ contains 36 atoms, and a $2 \times 6 \times 6$ k-point grid was used. The magnetic orderings of the experimental ground state were considered, namely ferromagnetic (FM) for SrCoO$_3$; antiferromagnetic (AFM) for LaFeO$_3$, LaCrO$_3$, and BM SrCoO$_{2.5}$.

The self-consistent Hubbard U (U$_{\rm LR}$) was calculated from first-principles by using the linear response approach proposed by Cococcioni and Gironcoli, in which U is determined by the difference between the screened and bare second derivative of the energy with respect to localized state occupations. The obtained U$_{\rm LR}$ values are 5.4 eV for Co in SrCoO$_3$, 4.2 eV for Ni in LaNiO$_3$, 5.6 eV for Fe in LaFeO$_3$, and 4.8 eV for Cr in LaCrO$_3$. For BM SrCoO$_{2.5}$, there are two types of Co sites, and the U$_{\rm LR}$ was determined to be 6.3 eV for the octahedral Co site and 4.9 eV for the tetrahedral Co site.

\subsection{\label{subsec:level1}QMC}


For this study, we utilized FN-DMC to obtain ground state properties of the selected family of ABO$_x$ materials.
We also calculated excited states for the set of insulating systems.
The high-performance QMC code, $\textsc{QMCPACK}$, was used for all FN-DMC calculations.
The QE software package was used throughout to generate the single-body orbitals that define the nodal hypersurface of the FN-DMC wave function.

To reach minimal fixed-node error, we variationally optimized the strength of a metal-site Hubbard interaction within LDA$+U$ for the full set of materials.
In the case of BM SrCoO$_{2.5}$, the $U$ value of the tetrahedral Co site was varied while keeping the difference between its value and the $U$ value at the octahedral site fixed at the difference predicted by linear reponse ($6.3-4.9\, {\rm eV} = 1.4\, {\rm eV}$).
For bulk SrCoO$_3$, we carefully verified that our LDA$+U$ orbitals were converged to the ground state by uniformly sampling the starting occupation matrix for the Co site within LDA$+U$.
We generated $36$ configurations, where all $10$ orbital occupations ($5$ within each spin-channel) were randomly chosen between $0$ and $1$ over a uniform distribution.
Furthermore, for SrCoO$_3$ we performed a more thorough nodal optimization by including scans over PBE$+U$ and PBE$0_w$ orbital sets, where line searches were performed over $U$ and the exact exchange fraction, $w$, of PBE$0$.

For the pseudopotentials, we employed the well-tested high-accuracy correlation-consistent ECPs (ccECPs) \cite{Bennett2017,Bennett2018,Annaberdiyev2018,Wang2019,annaberdiyev_accurate_2019}, where available.
In cases where ccECPs were not available, namely La and Sr, we generated the pseudopotentials as part of this work.
For the La and Sr pseudopotentials, we used a many-body testing protocol consistent with what was defined in the development of ccECPs to verify their accuracy.
The ccECPs Fe, Co, and Ni were softened to reach lower kinetic energy cutoff values and consequently lower computational cost.
To soften these atoms, their semi-local Gaussian parameter values were tuned to reduce the potential's amplitude at the origin in order to make them less plane-wave hard then were re-optimized within the full ccECP framework.
A cutoff of $730$ Ry was used for LaCrO$_3$ and all other materials used $670$ Ry -- leading to convergences of $0.05$--$0.06$ mHa/atom for all systems.
For self-consistency, an electronic convergence tolerance of $1\times10^{-10}$ Ry was used throughout.
For SrCoO$_3$, the SCF density was energetically converged to $0.01$ mHa/atom using a $12\times12\times12$ Monkhorst-Pack grid.
For all other systems, Monkhorst-Pack grids were used that were commensurate with the $12\times12\times12$ grid of SrCoO$_3$.

Experimental structures were taken from the ICSD database for the full set of materials.
The system-specific ICSD collection codes that were used are as follows: SrCoO$_3(77142)$, LaNiO$_3(173477)$, SrCoO$_{2.5}(162241)$, LaFeO$_3(184304)$, and LaCrO$_3(194033)$.
One-body finite-size effects were minimized by employing twist-averaged boundary conditions.
Fine supertwist grids with spacings of less than $0.15/\AA$ were used for all twist-averaged quantities.
Two-body finite size effects were partly reduced through the use of tiled supercells.
Supercells containing $8$ formula units were used for all the materials -- $40$ atoms in the case of SrCoO$_3$.
Cubic supercells were used when possible, namely, for SrCoO$_3$, LaFeO$_3$, and LaCrO$_3$.
For LaNiO$_3$ and BM SrCoO$_{2.5}$, we performed a brute force search over tiling matrix variations to find near-cubic supercells.


\begin{table}[h!]
\centering
\caption{Metal charges from radial density integration from DMC/LDA+$U_{\rm min}$. 
        $r_{\rm max}$ is taken to be the outer minima of the spherically averaged 
        total charge denstiy centered at the cation.}
\label{table_SI:qmc_metal_charges}
\begin{tabular}{l l l l l} 
 \hline\hline
       system    & atom  & $r_{\rm max}$ (\AA) & Charge (a.u.) & State (a.u.) \\
 \hline
    SrCoO$_3$    &    Co &    0.96 & 14.3513 & 2.6487  \\  
LaNiO$_3$(FM)    &    Ni &    1.02 & 16.2730 & 1.7270  \\
SrCoO$_{2.5}$    &   Co1 &    1.03 & 14.6202 & 2.3798  \\
SrCoO$_{2.5}$    &   Co2 &    1.03 & 14.7639 & 2.2361  \\
    LaFeO$_3$    &    Fe &    1.00 & 13.4973 & 2.5027  \\
    LaCrO$_3$    &    Cr &    0.99 & 11.2228 & 2.7772  \\
 \hline
\end{tabular}
\end{table}

\begin{table}[h!]
\centering
\caption{Metal charges from radial density integration from LDA+$U_{\rm min/DMC}$}
\label{table_SI:scf_metal_charges}
\begin{tabular}{l l l l l} 
 \hline\hline
	system    & atom  & rcut (\AA) & Charge (a.u.) & State (a.u) \\
 \hline
    SrCoO$_3$   &   Co  &       1.0  &      14.5941 & 2.4059     \\  
    LaNiO$_3$   &   Ni  &       1.0  &      16.1558 & 1.8061     \\
SrCoO$_{2.5}$   &  Co1  &       1.0  &      14.4792 & 2.5208     \\
SrCoO$_{2.5}$   &  Co2  &       1.0  &      14.6142 & 2.3858     \\
    LaFeO$_3$   &   Fe  &       1.0  &      13.5023 & 2.4977     \\
    LaCrO$_3$   &   Cr  &       1.0  &      11.2671 & 2.7329     \\
 \hline
\end{tabular}
\end{table}

\begin{figure*}[t]
\centering
\includegraphics[width=0.75\textwidth]{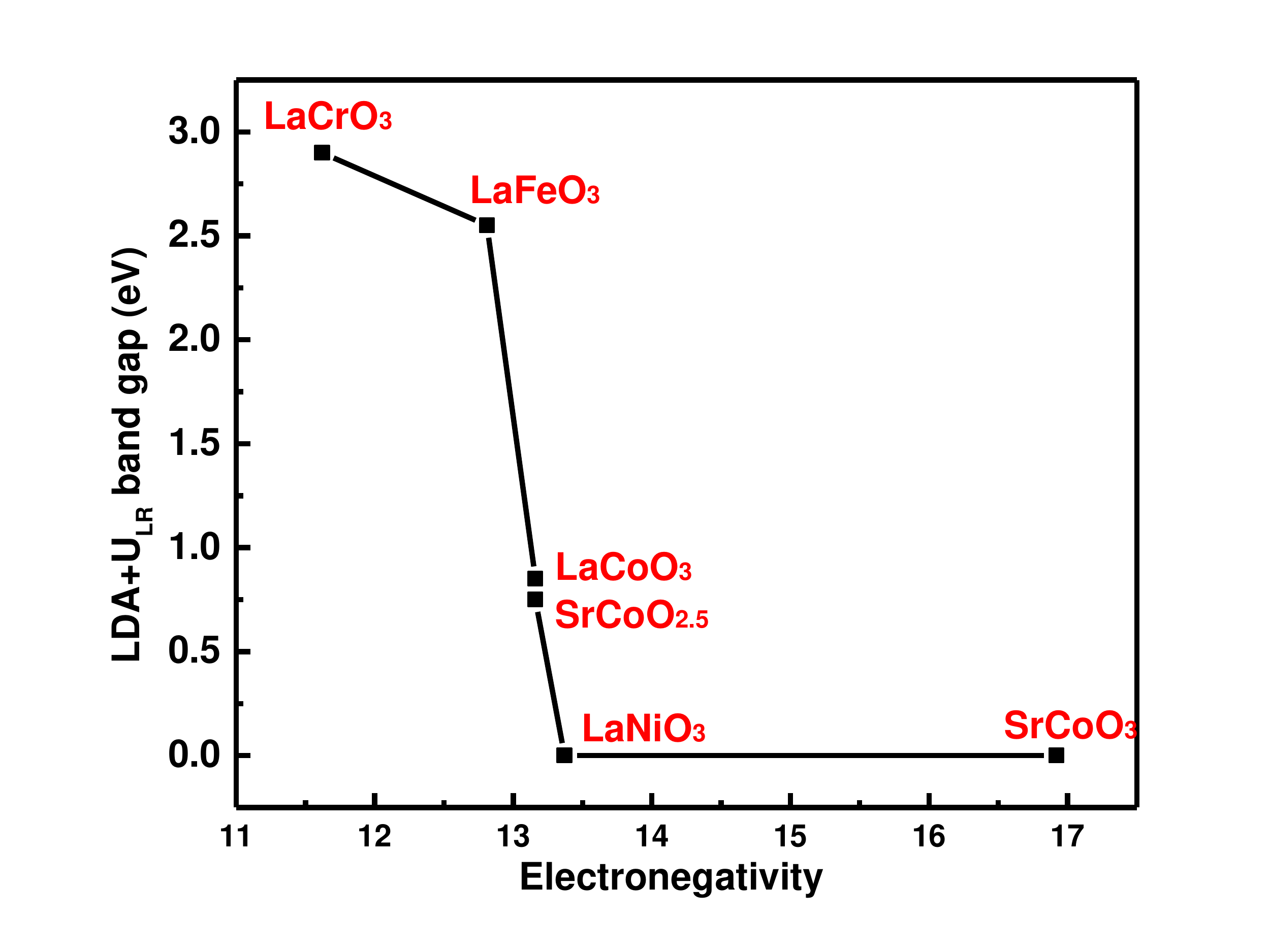}
\caption{Band gap of ABO$_x $as a function of the electronegativity of the B site cation.}
\label{fig_SI:FigX_GapVsElecNeg}
\end{figure*}

\begin{figure*}[t]
\centering
\includegraphics[width=\textwidth]{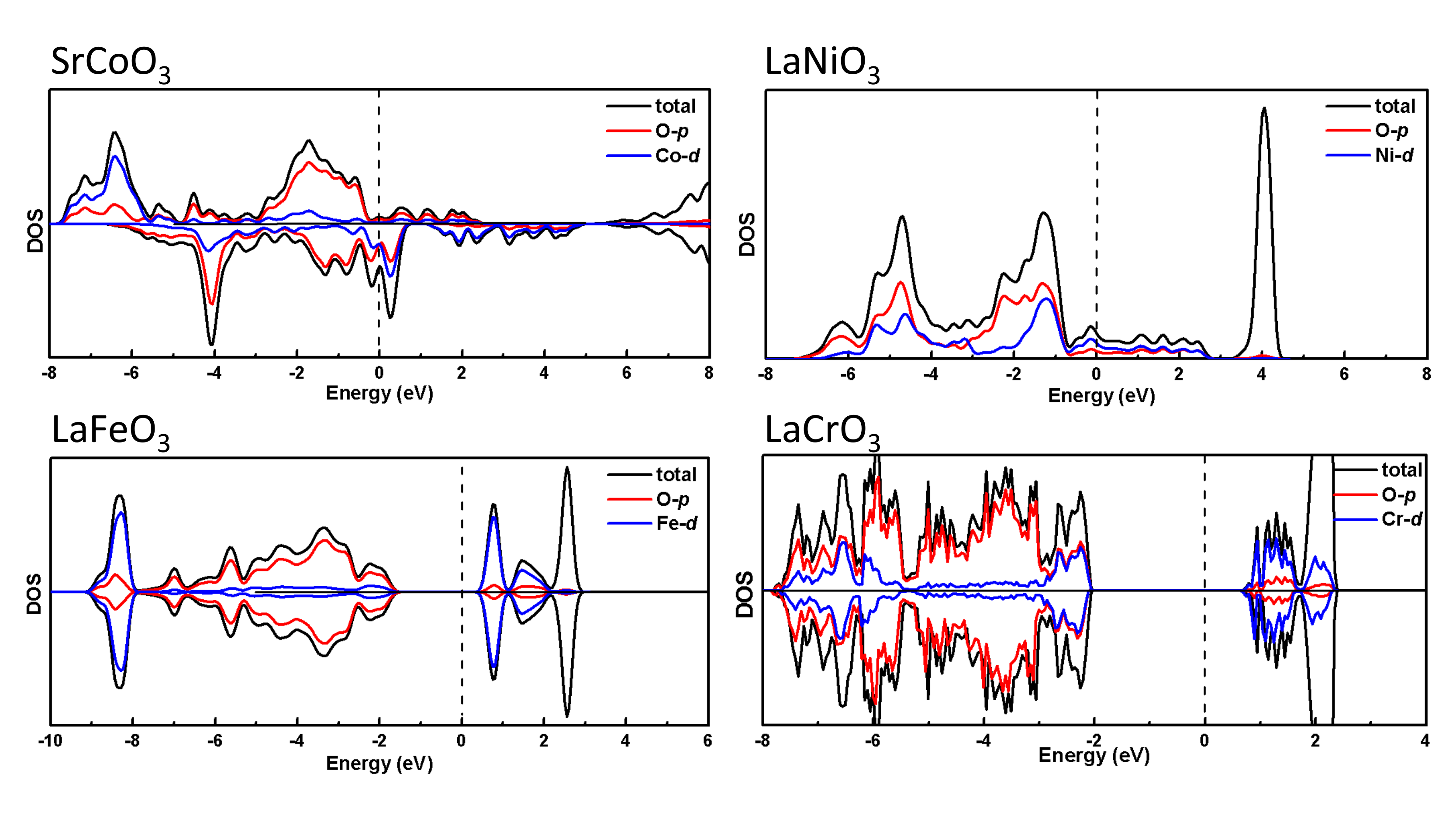}
\caption{DOS plots for SrCoO$_3, $LaNiO$_3, $LaFeO$_3, $and LaCrO$_3$. A gap-nature changing from p-p to p-d type can be observed.}
\label{fig_SI:FigS1_DOS}
\end{figure*}

\begin{figure*}[t]
\centering
\includegraphics[width=0.70\textwidth]{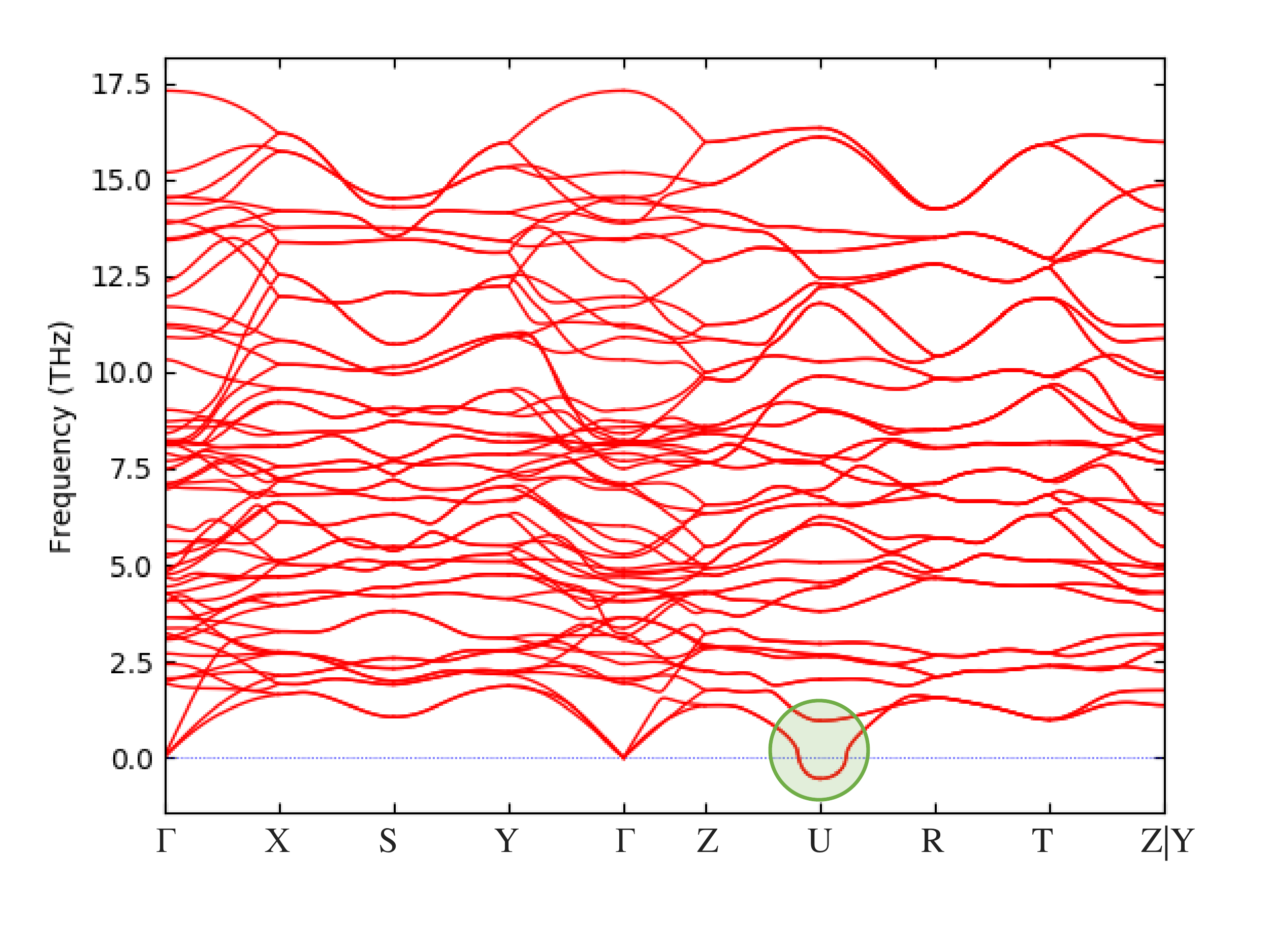}
\caption{Phonon dispersions of the hole-doped LaFeO$_3$.}
\label{fig_SI:FigY_HoleDopedPhonon_LFO}
\end{figure*}

\begin{figure*}[t]
\centering
\includegraphics[width=0.70\textwidth]{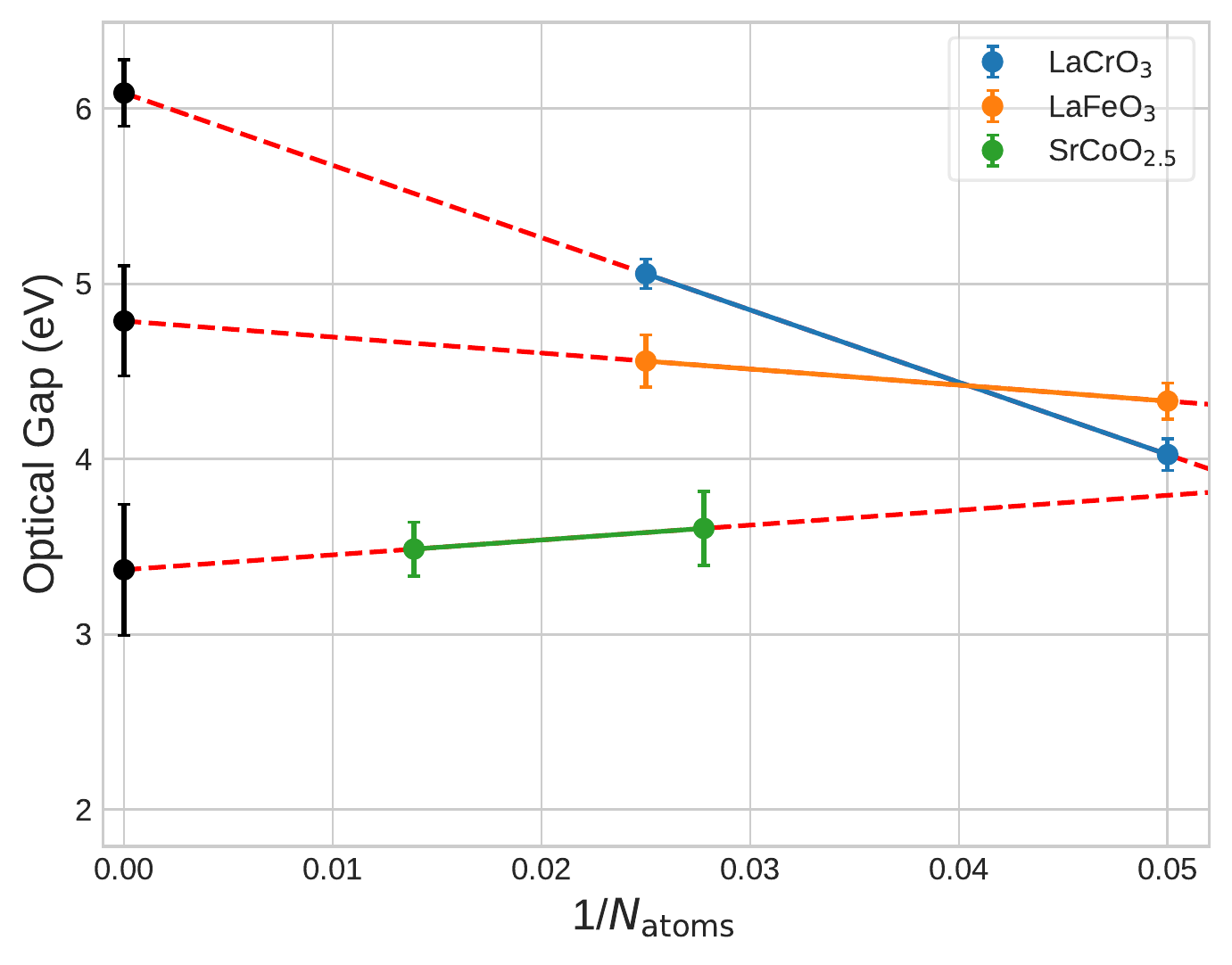}
        \caption{DMC optical gaps of LaCrO$_3$, LaFeO$_3$, and SrCoO$_{2.5}$ extrapolated to thermodynamic limit.}
\label{fig_SI:DMCBandGaps}
\end{figure*}

\begin{figure*}[t]
\centering
\includegraphics[width=0.70\textwidth]{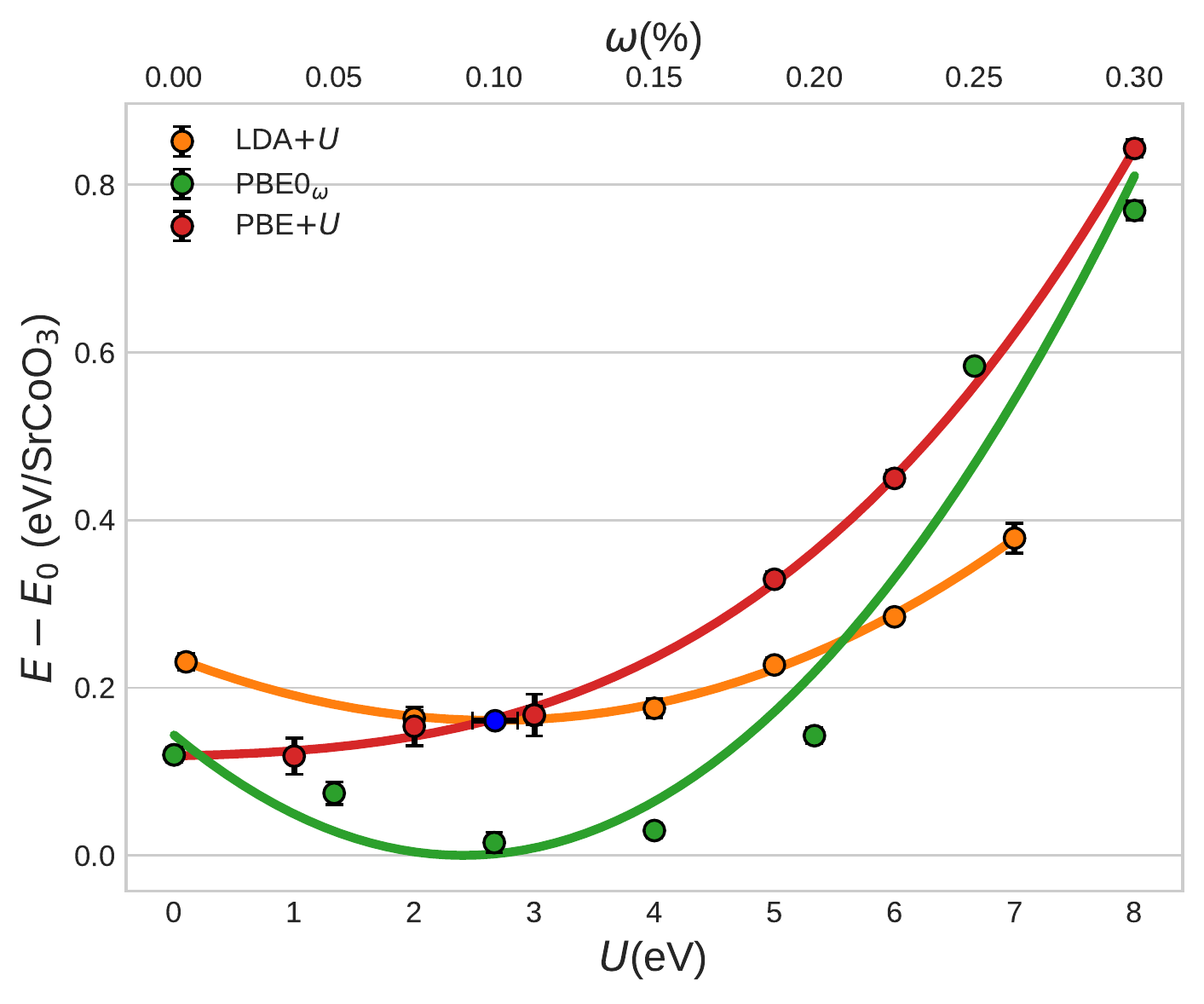}
        \caption{DMC/LDA+$U$ nodal optimization of SrCrO$_3$ using 40 atom simulation cell.}
\label{fig_SI:SCO_DMC_Nodal_Opt}
\end{figure*}

\begin{figure*}[t]
\centering
\includegraphics[width=0.70\textwidth]{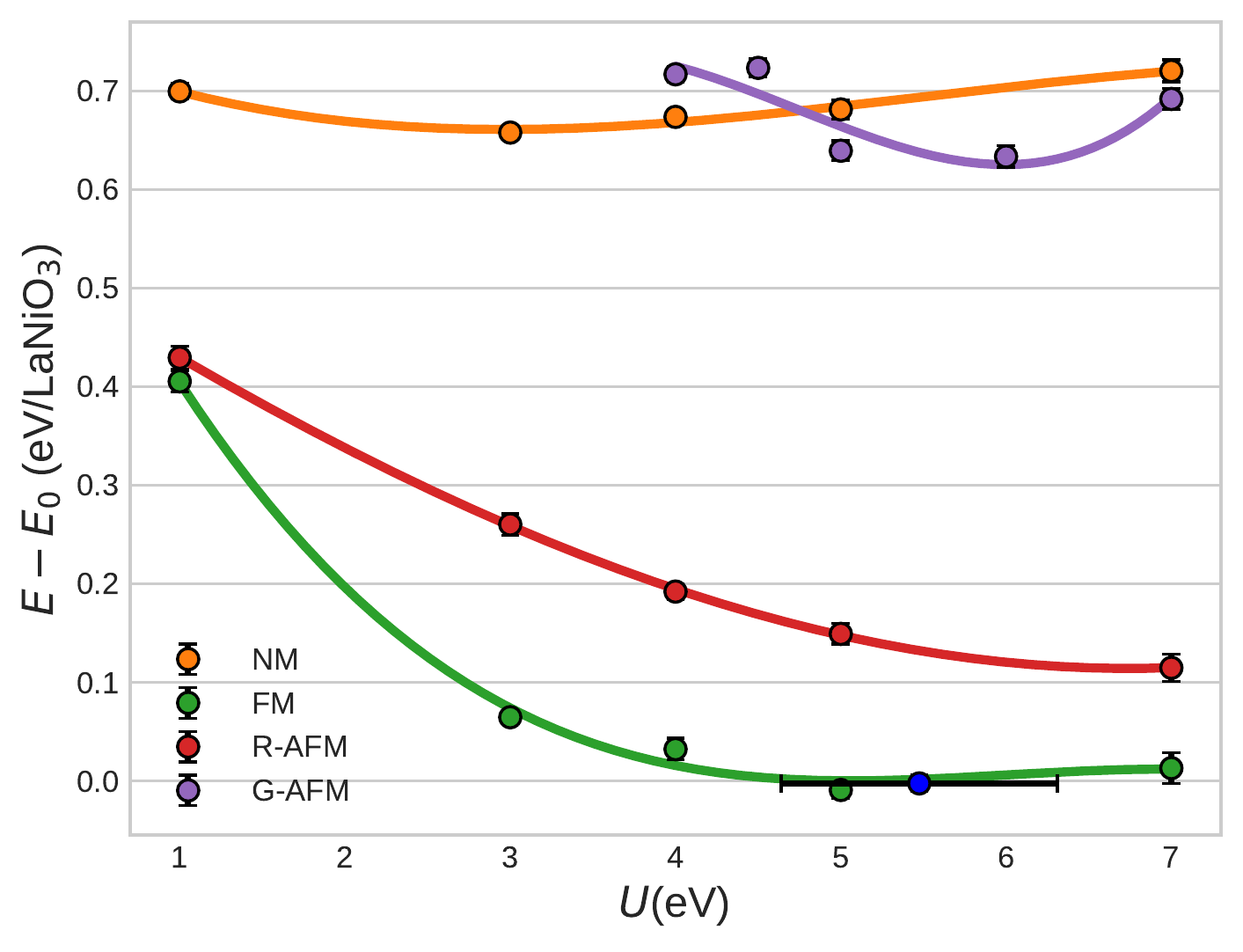}
        \caption{DMC/LDA+$U$ nodal optimization of LaNiO$_3$ using 40 atom simulation cell.}
\label{fig_SI:LNiO_DMC_Nodal_Opt}
\end{figure*}

\begin{figure*}[t]
\centering
\includegraphics[width=0.70\textwidth]{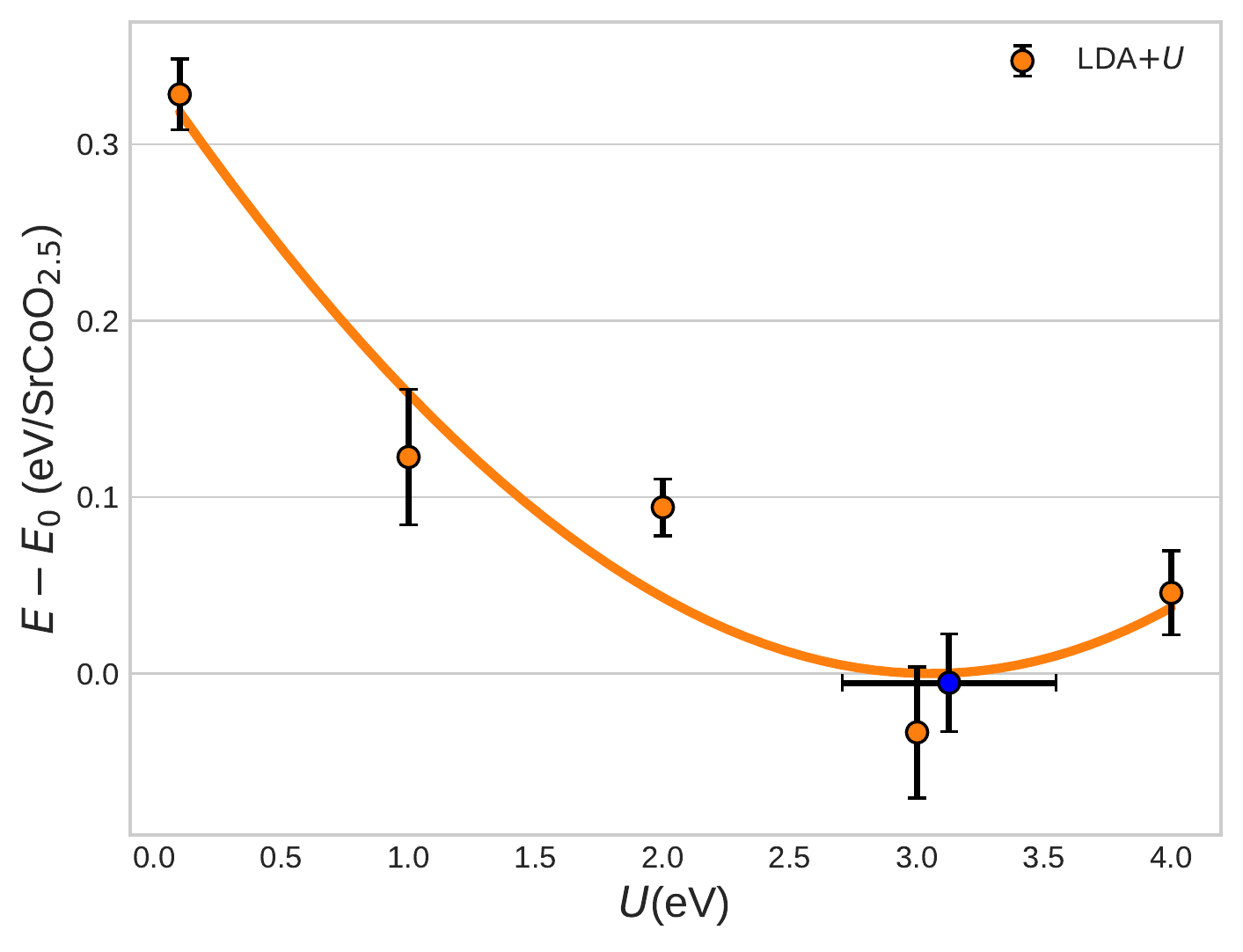}
        \caption{DMC/LDA+$U$ nodal optimization of SrCoO$_{2.5}$ using 80 atom simulation cell.}
\label{fig_SI:BM_SCO_DMC_Nodal_Opt}
\end{figure*}

\begin{figure*}[t]
\centering
\includegraphics[width=0.70\textwidth]{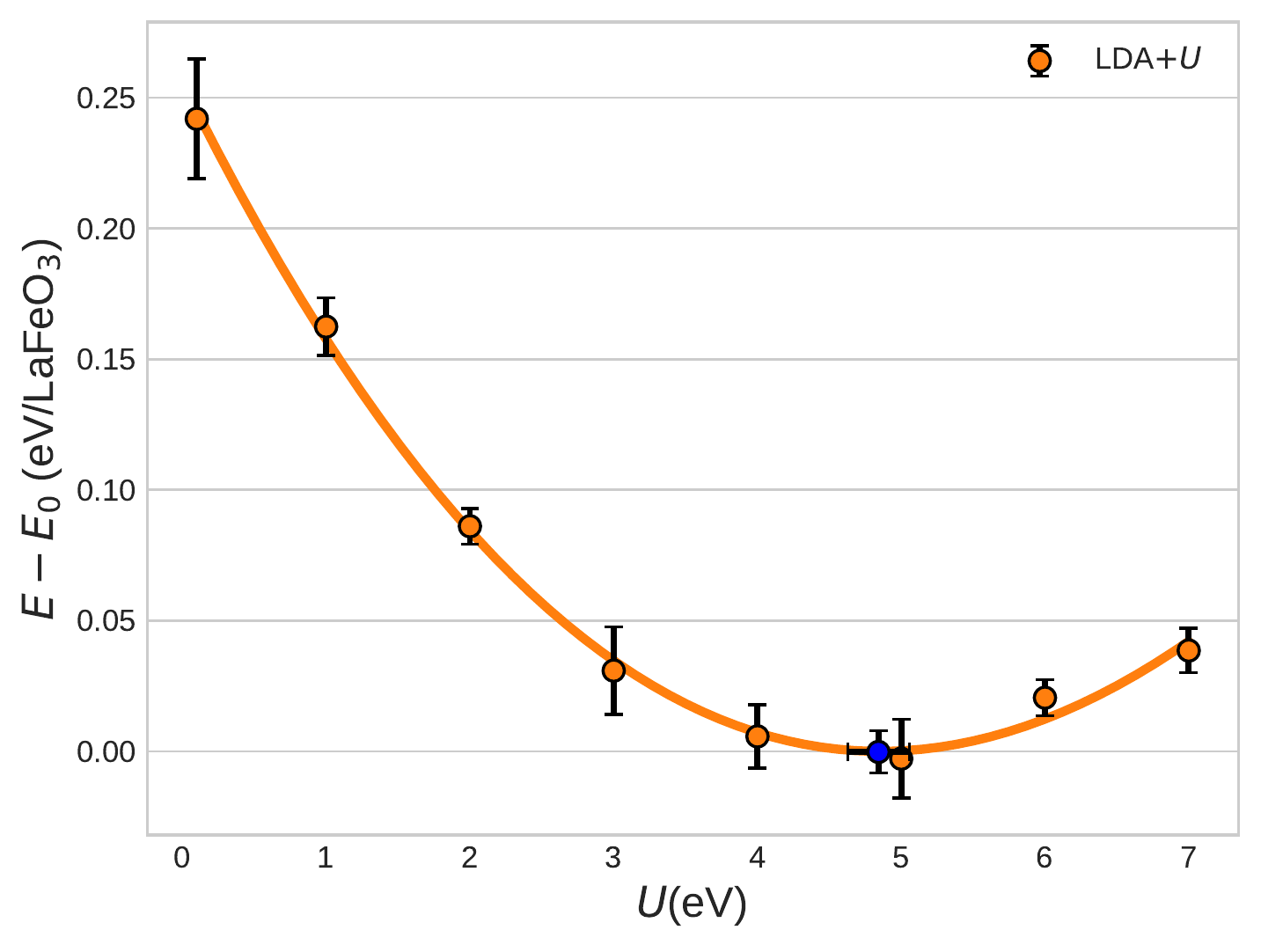}
        \caption{DMC/LDA+$U$ nodal optimization of LaCrO$_3$ using 40 atom simulation cell.}
\label{fig_SI:LCrO_DMC_Nodal_Opt}
\end{figure*}

\begin{figure*}[t]
\centering
\includegraphics[width=0.70\textwidth]{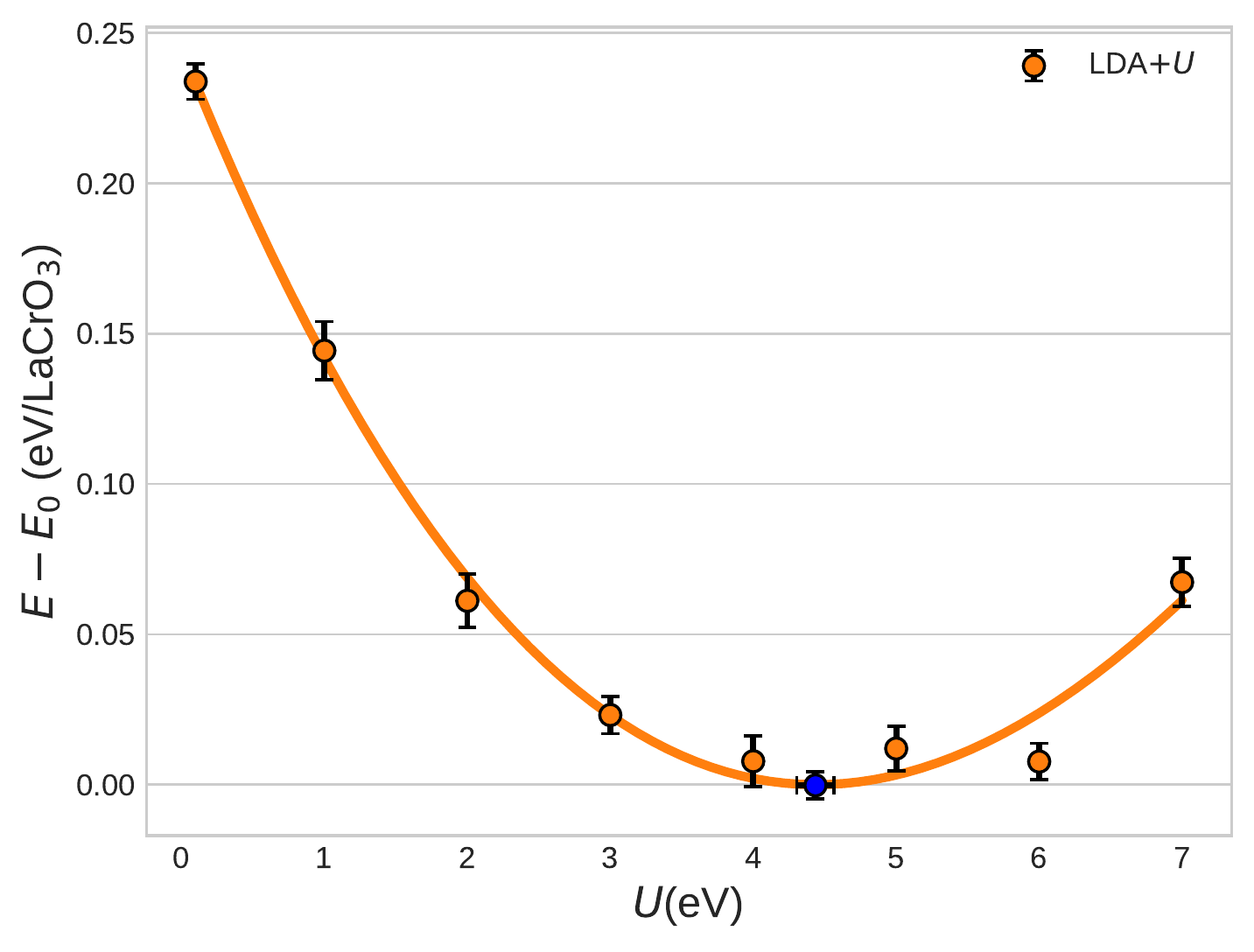}
        \caption{DMC/LDA+$U$ nodal optimization of LaFeO$_3$ using 40 atom simulation cell.}
\label{fig_SI:LFO_DMC_Nodal_Opt}
\end{figure*}

\begin{table}[h!]
\centering
\caption{$U$ values predicted from linear response ($U_{\rm LR}$) and FN-DMC nodal optimizations}
\label{table_SI:optimal_u_values}
\begin{tabular}{l l l l} 
 \hline\hline
      system    & $U_{\rm LR}$  & $U_{\rm min}$      \\
 \hline
    SrCoO$_3$   &    5.4.  &    2.7(2)     \\ 
LaNiO$_3$(FM)   &    4.2   &    5.5(8)     \\
SrCoO$_{2.5}$ (oct/tet) & 4.9,6.3  & 3.1(4),4.5(4) \\
    LaFeO$_3$   &   5.6     &  4.8(2) \\
    LaCrO$_3$   &   4.8     &  4.4(1) \\
 \hline
\end{tabular}
\end{table}

\begin{table}[h!]
\centering
\caption{Metal-oxygen bond lengths are listed for each system in \AA. The corresponding FN-DMC onsite oxygen charge is also listed in a.u.}
\label{table_SI:bond_lengths_and_charges}
\begin{tabular}{l l l l} 
 \hline\hline
      system    & Bond  & Bond Length & Oxygen Charge      \\
 \hline
    SrCoO$_3$   &   O-Co &  1.9175 &  6.4120\\ 
LaNiO$_3$(NM)   &  O1-Ni &  1.9378 &  6.4096\\ 
LaNiO$_3$(NM)   &  O2-Ni &  1.9378 &  6.4021\\ 
LaNiO$_3$(FM)   &  O1-Ni &  1.9378 &  6.4237\\ 
LaNiO$_3$(FM)   &  O2-Ni &  1.9378 &  6.4193\\ 
SrCoO$_{2.5}$   & O1-Co1 &  1.9407 &  6.4441\\ 
SrCoO$_{2.5}$   & O2-Co2 &  1.7900 &  6.3741\\ 
SrCoO$_{2.5}$   & O3-Co2 &  1.7947 &  6.3365\\ 
    LaFeO$_3$   &  O1-Fe &  2.0077 &  6.5845\\ 
    LaFeO$_3$   &  O2-Fe &  2.0000 &  6.4333\\ 
    LaCrO$_3$   &  O1-Cr &  1.9687 &  6.6497\\ 
    LaCrO$_3$   &  O2-Cr &  1.9634 &  6.4646\\ 
 \hline
\end{tabular}
\end{table}

\begin{figure*}[t]
\centering
\includegraphics[width=0.7\textwidth]{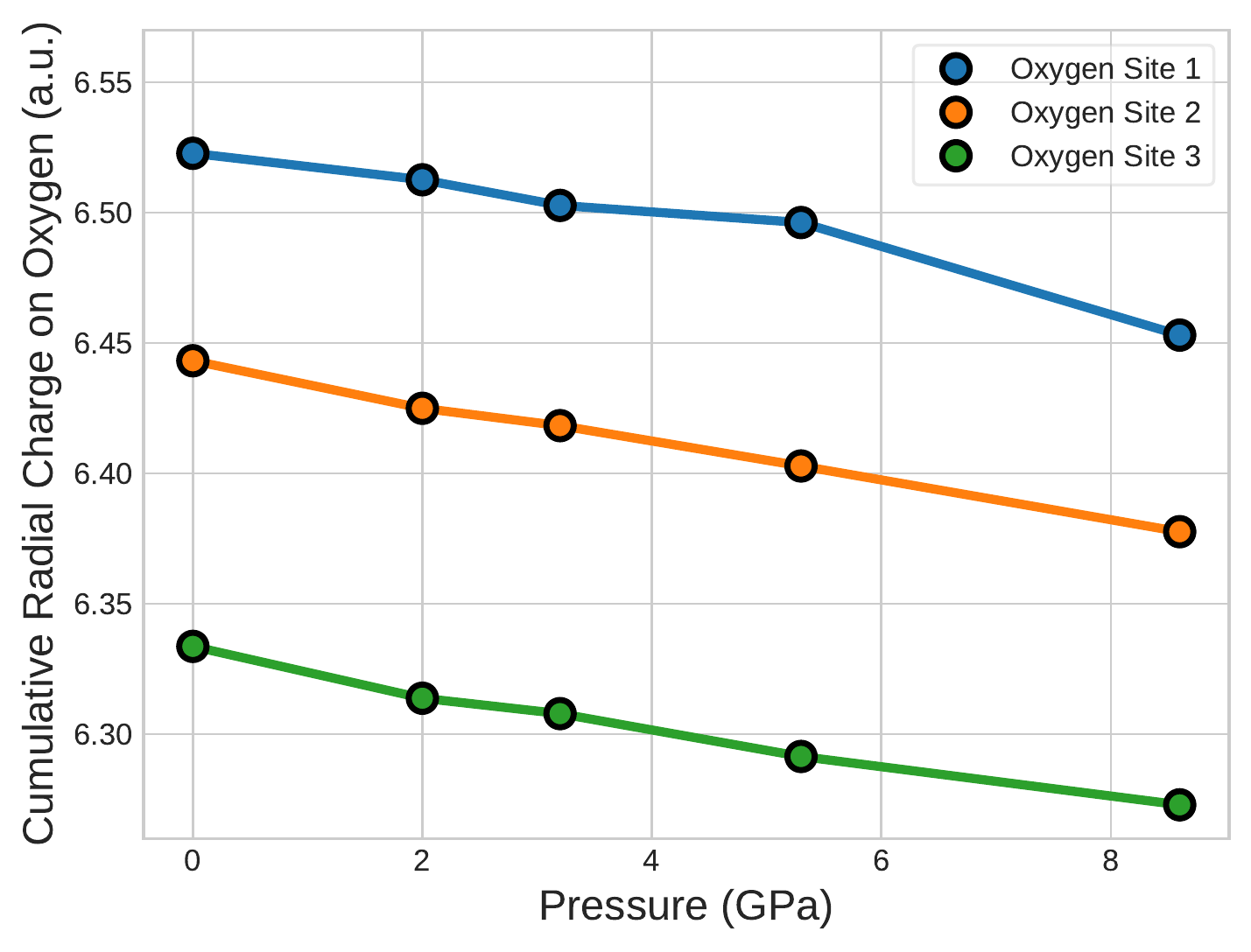}
\caption{}
\label{fig_SI:bm_sco_strained_o_populations}
\end{figure*}

\begin{table}[h!]
\centering
\caption{Oxygen charges from radial density integration from LDA+$U_{\rm min/DMC}$}
\label{table_SI:scf_oxygen_charges}
\begin{tabular}{l l l l} 
 \hline\hline
      system    & atom  & rcut (\AA) & Charge (a.u.) \\
 \hline
    SrCoO$_3$   &    O  &       1.1  &       6.32684 \\  
    LaNiO$_3$   &    O  &       1.1  &       6.31259 \\
SrCoO$_{2.5}$   &   O1  &       1.1  &       6.34444 \\
SrCoO$_{2.5}$   &   O2  &       1.1  &       6.23610 \\
SrCoO$_{2.5}$   &   O3  &       1.1  &       6.25541 \\
    LaFeO$_3$   &   O1  &       1.1  &       6.49029 \\
    LaFeO$_3$   &   O2  &       1.1  &       6.34051 \\
    LaCrO$_3$   &   O1  &       1.1  &       6.56356 \\
    LaCrO$_3$   &   O2  &       1.1  &       6.37763 \\
 \hline
\end{tabular}
\end{table}

\begin{table}[h!]
\centering
\caption{Oxygen charges from radial density integration from DMC/LDA+$U_{\rm min}$}
\label{table_SI:qmc_oxygen_charges}
\begin{tabular}{l l l l} 
 \hline\hline
      system    & atom  & rcut (\AA) & Charge (a.u.) \\
 \hline
    SrCoO$_3$  &     O &   1.1 &      6.41249 \\
    LaNiO$_3$  &    O1 &   1.1 &      6.40959 \\
    LaNiO$_3$  &    O2 &   1.1 &      6.40214 \\
SrCoO$_{2.5}$  &    O1 &   1.1 &      6.44413 \\
SrCoO$_{2.5}$  &    O2 &   1.1 &      6.37409 \\
SrCoO$_{2.5}$  &    O3 &   1.1 &      6.33649 \\
    LaFeO$_3$  &    O1 &   1.1 &      6.58540 \\
    LaFeO$_3$  &    O2 &   1.1 &      6.43631 \\
    LaCrO$_3$  &    O1 &   1.1 &      6.64961 \\
    LaCrO$_3$  &    O2 &   1.1 &      6.46568 \\
 \hline
\end{tabular}
\end{table}

\begin{table*}[h!]
\centering
\caption{On-site charges from LDA+U GBRV linear response}
\label{table_SI:scf_metal_charges_LR}
\begin{tabular}{l l l l l l l l l l l l} 
 \hline\hline
              &          & \multicolumn{5}{c}{L\"owdin-U$_{LR}$} \\ 
	system    & pressure & TM1 & TM2 & O1 & O2 & O3 \\ 
 \hline
    LaCrO$_3$   &   0    & 10.7008 & 12.4225 & 6.5860 & 6.5897 &   \\
    LaFeO$_3$   &   0    & 10.4151 & 15.238  & 6.4192 & 6.4227 &   \\
    LaNiO$_3$   &   0    & 10.4031 & 17.7272 & 6.2656 &    &   \\
    SrCoO$_3$   &   0    & 9.6326  & 16.5942 & 6.2383 &    &   \\  
SrCoO$_{2.5}$   &   0    & 9.6610  & 16.4557 & 6.3590 & 6.3409 & 6.3353 \\
SrCoO$_{2.5}$   &   2    & 9.6766  & 16.4644 & 6.3544 & 6.3344 & 6.3271 \\
SrCoO$_{2.5}$   &   3.2  & 9.6874  & 16.4692 & 6.3519 & 6.3306 & 6.3233 \\
SrCoO$_{2.5}$   &   5.2  & 9.6864  & 16.4764 & 6.3481 & 6.3257 & 6.3145 \\
SrCoO$_{2.5}$   &   8.8  & 9.6833  & 16.5029 & 6.3151 & 6.2977 & 6.3038 \\
 \hline
\end{tabular}
\end{table*}

\subsection{\label{subsec:level2}Cohesive Energies}

\begin{table}[h!]
  \centering
  \caption{Experimental cohesive energies}
  \label{table_SI:exp_cohesive_energies}
  \begin{tabular}{r r r r} 
    \hline
    \hline
    \multicolumn{4}{c}{LaCrO$_3$}\\
    \hline
    Reaction & T(K) & $\Delta_f H^{\circ}$(kJ/mol) &  $\Delta_f H^{\circ}$(eV) \\  
    \hline
    LaCrO$_3$(s)  & 298  &  -73.06   &  -0.757\footnotemark[1] \\
    Cr2O3(s)      & 0    &-1128.844  & -11.700\footnotemark[2] \\ 
    La2O3(s)      & 0    &-1787.36   & -18.524\footnotemark[3] \\ 
        O(g)      & 0    &  246.790  &   2.558\footnotemark[2] \\ 
       Cr(g)      & 0    &  395.340  &   4.097\footnotemark[2] \\ 
       La(g)      & 0    &  431.29   &   4.470\footnotemark[3] \\ 
       & & & \\
    \multicolumn{4}{c}{$E_{\rm coh}$ = 0.757+(11.7+18.524)/2+3(2.558)}\\
    \multicolumn{4}{c}{+4.097+4.47 eV = 32.11 eV}\\ 
    & & & \\
    \hline
    \multicolumn{4}{c}{LaFeO$_3$}\\
    \hline
    Reaction & T(K) & $\Delta_f H^{\circ}$(kJ/mol) &  $\Delta_f H^{\circ}$(eV) \\  
    \hline    
    LaFeO$_3$(s)  &  0  &  -64.58  & -0.669\footnotemark[1] \\ 
      Fe2O3(s)    &  0  & -819.025 & -8.489\footnotemark[2] \\ 
      La2O3(s)    &  0  &-1787.36  &-18.524\footnotemark[3] \\ 
          O(g)    &  0  &  246.790 &  2.558\footnotemark[2] \\ 
         Fe(g)    &  0  &  413.127 &  4.282\footnotemark[2] \\ 
         La(g)    &  0  &  431.29  &  4.470\footnotemark[3] \\ 
       & & & \\
    \multicolumn{4}{c}{$E_{\rm coh}$ = 0.669+(8.489+18.524)/2+3(2.558)}\\
    \multicolumn{4}{c}{+4.282+4.47 eV = 30.6 eV}\\ 
    & & & \\
    \hline
    \multicolumn{4}{c}{LaNiO$_3$}\\
    \hline
    Reaction & T(K) & $\Delta_f H^{\circ}$(kJ/mol) &  $\Delta_f H^{\circ}$(eV) \\  
    \hline    
   LaNiO$_3$(s)  &  1000 &  -46.07   & -0.477\footnotemark[1] \\ 
   Ni2O3(s)      &   298 & -489.5   & -5.073\footnotemark[5] \\ 
   La2O3(s)      &     0 &-1787.36   &-18.524\footnotemark[3] \\ 
         O(g)    &     0 &  246.790  &  2.558\footnotemark[2] \\ 
        Ni(g)    &     0 &  428.076  &  4.437\footnotemark[2] \\ 
        La(g)    &     0 &  431.29   &  4.470\footnotemark[3] \\ 
       & & & \\
    \multicolumn{4}{c}{$E_{\rm coh}$ = 0.477+5.073/2+18.524/2+3(2.558)}\\
    \multicolumn{4}{c}{+4.437+4.47 eV = 28.86 eV}\\ 
    & & & \\
 \hline
\end{tabular}
\footnotetext[1]{Reference \cite{cheng_navrotsky_zhou_anderson_2005}}
\footnotetext[2]{JANAF Tables}
\footnotetext[3]{https://data.nist.gov/od/id/mds2-2124}
\footnotetext[4]{Reference \cite{1954_Boyle_JAmerChemSoc_76_3835_Ni_and_Co_Oxides_Formation}}
\footnotetext[5]{Reference \cite{CRC_Handbook}}
\end{table}

\bibliography{si}